\documentclass[prb, reprint]{revtex4-2}
\usepackage{amsmath,amssymb}
\usepackage{bm}

\usepackage{mathtools}

\usepackage[dvipdfmx]{graphics}
\usepackage{dcolumn}
\usepackage[dvipsnames]{xcolor}
\usepackage[caption=false,position=top,font=normalsize]{subfig}
\usepackage{here}
\usepackage{bashful}

\usepackage{braket}
\usepackage[dvipdfmx, bookmarkstype=toc, colorlinks=true, linkcolor=cyan, pdfborder={0 0 0}, bookmarks=true, bookmarksnumbered=true]{hyperref}

\usepackage{hyperref}
\usepackage{tikz}
\usepackage{ulem}

\newcommand{\lr}[1]{\left({#1}\right)}

\newcommand{\lra}[1]{\left[{#1}\right]}
\newcommand{\lrb}[1]{\left\{{#1}\right\}}

\let\origlim\lim
\renewcommand{\lim}[2]{\origlim\limits_{{#1}\rightarrow{#2}}}

\newcommand{\Sum}[2]{\sum\limits_{#1}^{#2}}

\newcommand{\eps}{\epsilon}

\begin{document}

\title{Superconducting Diode Effect in Weak Localization Regime}

\author{Naratip Nunchot}
\altaffiliation{Present address: Department of Applied Physics, Graduate School of Engineering, The University of Tokyo, Hongo, Tokyo 113-8656, Japan}
\affiliation{Department of Physics, Graduate School of Science, Kyoto University, Kyoto 606-8502, Japan}

\author{Youichi Yanase}
\affiliation{Department of Physics, Graduate School of Science, Kyoto University, Kyoto 606-8502, Japan}

\date{\today}

\begin{abstract}

We study a dirty two-dimensional superconductor with Rashba spin-orbit coupling and in-plane Zeeman fields described by the nonlinear sigma model that includes the Cooper and long-range Coulomb interactions. The renormalized Ginzburg-Landau theory, which includes the weak localization effects at the one-loop level, is constructed using the Keldysh functional formalism. It is shown that the transition temperature and magnetic field, as well as the tricritical point appearing in the phase diagram, are suppressed by the interactions. Nevertheless, we have found a universal behavior in the high-transition-temperature regime that demonstrates the robustness of the superconducting diode effect against the interactions. The conductivity of the resistive states emerging after the superconducting states are destroyed by the critical current is also calculated, and localization behaviors are demonstrated.

\end{abstract}

\maketitle 

\section{Introduction} 
The phenomenon in which the critical currents of superconductors become nonreciprocal, known as \textit{the superconducting diode (SD) effect}~\cite{Ando, Bauriedl}, has attracted much attention. Theoretical studies have been carried out for various types of noncentrosymmetric superconductors~\cite{Yuan, Daido1, He-Nagaosa, Aoyama, Nunchot1, Banerjee, Nunchot2}. Not only clean systems but also disordered systems have been studied at the mean-field (MF) level~\cite{Ilic1, Ikeda2022, Hasan}. If electron systems are disordered enough, they are well described by the nonlinear sigma model (NLSM)~\cite{Wegner, Hikami, Finkelstein, Kamenev1}, and dirty superconductors can also be described by the NLSM~\cite{Feigelman, Yurkevich, Levchenko1, Konig}. Recently, an NLSM for a dirty superconductor with spin-orbit coupling (SOC) and Zeeman field was derived \cite{Virtanen} and used to investigate the SD effect in Josephson junctions at the MF level~\cite{Ilic2}. However, the \textit{Anderson localization effect}, a correction to the MF theory in dirty systems, has not yet been highlighted in studies of the SD effect.

The role of Anderson localization in superconductors is an important topic of ongoing debate. Without electron-electron (e-e) interactions, the properties of dirty $s$-wave superconductors at zero magnetic field remain unchanged with respect to the strength of non-magnetic disorder in the weak localization (WL) regime, known as the Anderson theorem \cite{Anderson}. However, if the long-range Coulomb interaction exists, the transition temperature is largely suppressed even in the WL regime \cite{Maekawa, Finkelstein1, Oreg}. Moreover, the transition temperature can also be enhanced by the multifractality effect~\cite{Feigelmanloffe, Burmistrov1, Burmistrov11, Burmistrov2}, where the short-range nature of the e-e interactions is important. It was also shown that, in the WL regime, the Coulomb interaction lowers the critical current of conventional dirty superconductors at zero magnetic field and low temperatures~\cite{Fukuyama, Takayanagi}.

In this work, we study the role of WL correction from the Cooper and long-range Coulomb interactions, which are ubiquitous in the solid state systems \cite{Raffy1983, Graybeal1984, Chand2012, Noat2013, Banerjee2017}, on the SD effect. In normal metals, the localization behavior can be changed by the SOC. It is thus interesting to consider whether the WL behavior of the resistive states that emerge after the superconducting (SC) states are destroyed by the critical currents can be enhanced while maintaining the high efficiency of the SD effect. If this is the case, the resistive states may approach the Anderson insulating phase as the disorder strength increases. This would enable the realization of an ideal superconducting diode with no heat loss that can switch between a superconducting state with current flow and an insulating state with no current flow. To explore this possibility, we investigate the relationship between the efficiency of the SD effect in SC states and the WL correction to conductivity in resistive states arising when the current surpasses the critical current. We work with the NLSM for a dirty two-dimensional (2D) noncentrosymmetric superconductor, and establish a renormalized Ginzburg-Landau (GL) theory at the one-loop level. While many studies \cite{Burmistrov2, Andriyakhina1, Nosov, Andriyakhina2} have been conducted in the replica formalism, we formulate the theory with the Keldysh functional formalism~\cite{Kamenev1, Feigelman, Levchenko1, Schwiete1}. This enables us to use the Keldysh action of the NLSM for the spin-orbit coupled system \cite{Virtanen} as the basis for our analysis.

The rest of this work is organized as follows. First, we introduce the model in Sec. \ref{Model}. Some notations and detailed calculations are described in Appendix \hyperref[AppendixA]{A}. In Sec. \ref{Method}, we present our method of using the Keldysh functional formalism to obtain the renormalized GL free energy, which is the central object for determining the SC states and calculating the supercurrent. Some lengthy equations and their derivations are shown in Appendices \hyperref[AppendixB]{B}, \hyperref[AppendixC]{C}, and \hyperref[AppendixD]{D}. In Sec. \ref{Numerical}, we present the numerical results on magnetic field-temperature phase diagrams and the efficiency of the SD effect. The latter is related to the WL conductivity of the normal states that emerge after the currents break the SC states closely, to conclude a crucial relationship between them. Other minor results in this section can be found in Appendices \hyperref[AppendixE]{E}, \hyperref[AppendixF]{F}, and \hyperref[AppendixG]{G}. We end this paper in Sec. \ref{Summary}.

\section{Model}\label{Model} 
We consider a dirty 2D superconductor aligned in the $xy$-plane with the Rashba SOC and a Zeeman field $h$ in the $x$-direction. It was shown in \cite{Virtanen} that in the Keldysh functional formalism, the partition function of the system without e-e interactions is given as $Z=\int D[Q,\bm{\Delta}] \exp\,(i\tilde{S}[Q,\bm{\Delta}])$, where 

\begin{equation}
\tilde{S}[\bm{\Delta},Q] = \tilde{S}_{\rm HS }[\bm{\Delta}] + \tilde{S}_{\rm N}[\bm{\Delta},Q] + S_{\rm T}[Q]. \label{SS1}
\end{equation}
Here, $\bm{\Delta}\equiv \bm{\Delta}(\bm{r},t)$ is the general gap function depending on space $\bm{r}$ and time $t$. $Q$ is the $8\times8$ matrix field which satisfies $Q^2 = 1$ and $\mathrm{Tr} \, \{Q\}=0$, where the trace $\mathrm{Tr}\,\{\cdot\}$ is performed with respect to space, time, and matrix elements. $\tilde{S}_{\rm HS}$ is responsible for the Hubbard-Stratonovich transformation associated with the SC state. $\tilde{S}_{\rm N}$ and $S_{\rm T}$ are the non-interacting and topological parts, respectively. $S_{\rm T}$ also serves as the origin of the SD effect \cite{Ilic2}. To proceed further, let $\gamma_{\mu},~\tau_{\mu}$, and $\sigma_{\mu}$, where $\mu =0,\dots,4$, denote the Pauli matrices in the Keldysh, Nambu, and spin spaces, respectively. We also write $\gamma_{\rm cl}=\gamma_0, \gamma_{\rm q}=\gamma_1$, and often abbreviate $\gamma_{\rm cl}$ for convenience. Moreover, the terms in the form $\gamma_{\mu}\check{A}$ stand for the expression $\gamma_{\mu}\otimes \check{A}$, where $\otimes$ is a tensor product and the check symbol $\check{(\cdot)}$ highlights that the objects are $4\times4$ matrices. Using these notations, the explicit forms of the terms in Eq. (\ref{SS1}) are expressed as follows:
\begin{align}  
\tilde{S}_{\rm HS}&= 4\nu \Gamma_{\rm c}^{-1} \int dtd\bm{r} \,  \lr{ \bm{\Delta}_{\rm cl}\bm{\Delta}^*_{\rm q } +  \bm{\Delta}^*_{\rm cl}\bm{\Delta}_{\rm q } }, \label{SGp}  \\
\tilde{S}_{\rm N}&= \frac{i\pi \nu}{8} \mathrm{Tr}\lrb{ D(\gamma_{\rm cl}\check{\nabla}_i Q)^2 + 4i \bm{\Omega} Q   }, \label{Snr} \\
S_{\rm T} &=  -\frac{i\pi\nu}{8}\frac{D\tau}{m} \mathrm{Tr} \lrb{ \gamma_{\rm cl} \check{F}_{xy} \, Q [\gamma_{\rm cl}\check{\nabla}_{x}Q,\gamma_{\rm cl}\check{\nabla}_{y} Q]   }. \label{Sp}
\end{align}
Here, $\bm{\Omega} = \gamma_{\rm cl} \check{E} + \gamma_{\rm cl}\check{\bm{\Delta}}_{\rm cl} + \gamma_{\rm q}\check{\bm{\Delta}}_{\rm q}$ and $\check{\bm{\Delta}}_{\rm cl(q)} = \tau_+\bm{\Delta}_{\rm cl(q)}-\tau_-\bm{\Delta}^*_{\rm cl(q)}$ with $\tau_{\pm} = (\tau_1\pm i\tau_2)/2$. Using $\check{\bm{\Delta}}_{\rm cl(q)}$, the general gap function $\bm{\Delta}$ in the integration measure of $Z$ is expressed as $\bm{\Delta}=\gamma_{\rm cl}\check{\bm{\Delta}}_{\rm cl} + \gamma_{\rm q}\check{\bm{\Delta}}_{\rm q}$. The integration $\int dt d\bm{r}$ is performed over a finite spatial area $\mathcal{A}$ and a long time interval $\mathcal{T}$, where the limit $\mathcal{T}\rightarrow\infty$ is taken when deriving the modified Usadel equation and gap equation, as well as when calculating conductivity. $\check{\nabla}_i$ is a covariant derivative defined as $\check{\nabla}_i \check{G} \equiv \nabla_i \check{G} - i[\check{A}_i, \check{G}]$, where $\check{A}_i$ is a matrix gauge field given as $\check{A}_x =\check{L}_{\rm r}(\tau_3 A_x + \sigma_2\alpha) \check{L}_{\rm r} = \tau_3A_x - \sigma_2\alpha$ and $\check{A}_y =\check{L}_{\rm r}(\tau_3 A_y - \sigma_1\alpha) \check{L}_{\rm r} =\tau_3 A_y-\sigma_3\alpha$ with $\check{L}_{\rm r}=\mathrm{diag} \, \{\hat{L}_{\rm r},\hat{L}_{\rm r}\}$ and $\hat{L}_{\rm r} = (\sigma_1+\sigma_3)/\sqrt{2}$. Here, $A_i$ is an ordinary gauge field, and the terms proportional to the Rashba SOC constant $\alpha$ are often referred to as spin gauge fields. To be consistent with the helical SC state discussed below, we let the gauge field $A_i$ be uniform from now on. Using $\check{L}_{\rm r}$,  $\check{A}_x$, and $\check{A}_y$, we have $\check{E}=\check{L}_{\rm r}(\eps\tau_3 + h\sigma_1)\check{L}_{\rm r} = \tau_3(\eps + h\sigma_3)$, where $\eps$ is energy, and $\check{F}_{xy} = \partial_x\check{A}_y-\partial_y\check{A}_x-i[\check{A}_x,\check{A}_y]= 2\alpha^2\tau_0\sigma_1$. Note that from the explicit form of $\check{F}$, $S_{\rm T}$ is proportional to the D'yakonov-Perel (DP) relaxation rate $\Gamma_{\rm r} = 2D\alpha^2$ where $D=v^2_{\rm F}\tau/2$ is the 2D diffusion constant, $v_{\rm F}$ is the Fermi velocity, and $\tau$ is the scattering time. The remaining constants $\nu=m/2\pi$ and $\Gamma_{\rm c}$ denote the 2D density of states \textit{per spin} and the SC coupling constant, respectively. The matrix forms of $\check{A}$, $\check{E}$, and $\check{F}$, the explicit calculations used to derive them, and the explicit form of the trace $\mathrm{Tr}$ can be found in Appendix \hyperref[AppendixA]{A}.

In the following, we consider only the static helical SC state, namely $\bm{\Delta}^{\rm Helical}_{\rm cl(q)}(\bm{r},t)\equiv\Delta_{\rm cl(q)}\exp\,(i\bm{P}\cdot\bm{r})$, where $\bm{P}$ is a Cooper-pair momentum and $\Delta_{\rm cl (q)}\in\mathbb{R}$ is the classical (quantum) component of the SC gap which is uniform in space and time. Note that the factor $\exp\,(i\bm{P}\cdot\bm{r})$ can be cancelled out by a U(1) transformation. Performing the transformation results in the shift of gauge field as $A_i\rightarrow A_i+P_i/2$. Since we set the Zeeman field $h$ to point out in the $x$-direction, the SD effect occurs when a supercurrent flows in the $y$-direction. We thus set $A_x=P_x=0$ and rewrite $A_y+P_y/2$ as $A_y$. Moreover, as the general form of the gap function is included in Eqs. (\ref{SS1}), (\ref{SGp}), and (\ref{Snr}), the fluctuation around the helical states must be integrated out. As a result, the Cooper interaction $S^{(\rm c)}_{\rm int}$ appears, and only the field $Q$ and the static uniform SC gaps, namely $\Delta$, remain in the integral measure of the partition function $Z=\int D[Q,\Delta]\exp\,(iS[Q,\Delta]+iS^{\rm(c)}_{\rm int}[Q])$, where 
\begin{align}  
S[\Delta,Q] = S_{\rm HS }[\Delta] + S_{\rm N}[\Delta,Q] + S_{\rm T}[Q]. \label{S1} 
\end{align}
While $S_{\rm T}$ remains the same form as Eq. (\ref{Sp}), the others are expressed as
\begin{align}
S_{\rm HS} &= 8\nu \Gamma_{\rm c}^{-1} \int dtd\bm{r} \,  \lr{ \Delta_{\rm cl}\Delta_{\rm q } },  \label{Sgp2} \\
S_{\rm N} &= \frac{i\pi \nu}{8} \mathrm{Tr}\lrb{ D(\gamma_{\rm cl}\check{\nabla}_i Q)^2 + 4i \Omega Q },  \label{Snr3}\\
\Omega &= \gamma_{\rm cl} \check{E} + \gamma_{\rm cl}\check{\Delta}_{\rm cl} + \gamma_{\rm q}\check{\Delta}_{\rm q}, \label{Esu}
\end{align}
and $\check{\Delta}_{\rm cl(q)} = i\tau_2\Delta_{\rm cl(q)}$. The general gap function $\bm{\Delta}$ in the integration measure of $Z$ also becomes $\Delta = \gamma_{\rm cl}\check{\Delta}_{\rm cl} + \gamma_{\rm q}\check{\Delta}_{\rm q}$. For the Cooper interaction term $S^{\rm (c)}_{\rm int}$,\begin{align}
S^{\rm (c)}_{\rm int} \!= \!\tilde{\Gamma}_{\rm c} \!  \Sum{i=1,2}{}  \int \! dtd\bm{r} \,  \mathrm{tr} \! \lrb{\gamma_{\rm cl} \tau_i  Q_{tt}(\bm{r})} \! \mathrm{tr} \! \lrb{\gamma_{\rm q} \tau_i  Q_{tt}(\bm{r})}, \label{Cooper_int}
\end{align}
where $\tilde{\Gamma}_{\rm c} = -\pi^2\nu\Gamma_{\rm c}/16$ and $\mathrm{tr}\,\{\cdots\}$ denotes a trace only in matrix elements. The comparison of Eq. (\ref{Cooper_int}) with the previous work and the detailed calculations brought to Eqs. (\ref{Sgp2}), (\ref{Snr3}), and (\ref{Cooper_int}), which involve performing the U(1) transformation and integrating the gap fluctuation, as shown in Appendix \hyperref[AppendixA]{A}.

In addition to the Cooper interaction, we take into account the Coulomb interaction $S^{\rm (s)}_{\rm int}[Q]$. Following the procedure in Ref.~\cite{Schwiete1}, we came to the expression
\begin{align}    
S^{\rm (s)}_{\rm int} = \tilde{\Gamma}_{\rm s}  \int dtd\bm{r} \, \mathrm{tr} \lrb{\gamma_{\rm cl} \tau_0 Q_{tt}(\bm{r})} \mathrm{tr}\lrb{\gamma_{\rm q} \tau_0 Q_{tt}(\bm{r})},   \label{Coulomb_int}     
\end{align}
where $\tilde{\Gamma}_{\rm s} = -\pi^2\nu\Gamma_{\rm s}/16$ and $\Gamma_{\rm s}=1$ in the long-range limit. As mentioned in the introduction, we consider only the long-range limit of the Coulomb interaction. In the following, the analysis is based on the total partition function $Z_{\rm tot}=\int D[\Delta,Q]\exp\,(iS_{\rm tot}[\Delta,Q])$, where 
\begin{equation}
S_{\rm tot}[\Delta, Q] = S[\Delta,Q] + S^{\rm (c)}_{\rm int}[Q] + S^{\rm (s)}_{\rm int}[Q]. \label{SQ_tot}
\end{equation}

Despite being beyond our scope, it is instructive to note that the triplet channel of e-e interactions can be neglected in any case if the triplet diffusive modes acquire a large mass. The diffusive modes, which are cooperon and diffuson modes, will be described in Sec. \ref{MethodB}. In our case, the mass of the triplet diffusive modes is the DP relaxation rate $\Gamma_{\rm r}$. Further details are discussed in Appendix \hyperref[AppendixD5]{D5}.

\section{Method}
\label{Method}

In the NLSM framework, the system is described at the MF level by searching the saddle matrix field $Q=\bar{Q}$ that optimizes the action under the constraints $\bar{Q}^2=1$ and $\mathrm{Tr} \, \{\bar{Q}\}=0$. It is possible to optimize the total action $S_{\rm tot}$ to obtain an MF theory that includes the contribution of e-e interactions \cite{Burmistrov2}. However, the corresponding Usadel equation has a nonlinear form that is difficult to solve. We thus take an alternative track \cite{Burmistrov2}: Setting a saddle point and integrating out fluctuations around it. This ultimately leads to an effective action of the SC gap which includes both the SD and WL effects as perturbative parts. The procedure to achieve such effective action is outlined below. 

\subsection{Effective action} 
\label{MethodA}

We consider the Gaussian fluctuations around an SC saddle point $\bar{Q}_{\rm s}$ whose form will be specified later. Although it is possible to adopt the normal saddle point $\bar{Q}_{\rm N}$ as the origin of expansion, the high-order calculation of perturbation theory is needed \cite{Levchenko1}. Time and effort can be saved by considering the SC saddle point \cite{Yurkevich}. Since $\bar{Q}_{\rm N}$ and $\bar{Q}_{\rm s}$ satisfy $\bar{Q}^2_{\rm N} = \bar{Q}^2_{\rm s}=1$, there is a matrix $U$ such that $\bar{Q}_{\rm s} = U\bar{Q}_{\rm N} U^{-1}$. The saddle point configuration is the same as the Keldysh fermionic action \cite{Kamenev1}. That is $\bar{Q} = U_{\rm K}\Lambda U_{\rm K}$, where $\Lambda = \mathrm{diag}\,\{\check{\Lambda}_{\rm R},\check{\Lambda}_{\rm A} \}$ and $\rm R(A)$ denotes the retarded (advanced) component. In the energy representation, $U_{\rm K}$ is 
\begin{align} 
U_{\mathrm{K}\eps} & = \begin{pmatrix}
\tau_0 & F_{\eps}\tau_0 \\
0 & -\tau_0 \end{pmatrix},  \label{Ukeys}
\end{align} 
where distribution in thermal equilibrium leads to $F_{\eps} = \tanh \eps/(2T)$ with temperature $T$. Using the above equations, we can rewrite $\bar{Q}_{\rm N(s)} = U_{\rm K}\Lambda_{\rm N(s)} U_{\rm K}$, where $\Lambda_{\rm N(s)}=\mathrm{diag}\,\{  \check{\Lambda}_{\rm N(s),R}, \check{\Lambda}_{\rm N(s),A}   \}$. The form of $\Lambda_{\rm N}$ in the energy representation is $\Lambda_{\mathrm{N}\eps} = \mathrm{diag}\lrb{ \tau_3,-\tau_3}$ \cite{Kamenev1, Feigelman}. On the other hand, $\Lambda_{\rm s}$ is generally a function of the \textit{spectral angle} $\theta$, which will be specified later. Note that $U$ is also a function of $\theta$, just as $\Lambda_{\rm s}$ is.

 Next, we perform the transformation $Q = U \tilde{Q} U^{-1}$ and then the transformation $\tilde{Q} =\exp\lr{W/2} \Lambda_{\rm N} \exp\lr{-W/2}$. Thus, $Q$ can be expanded to the quadratic order of $W$ as $\tilde{Q} = \Lambda_{\rm N} + \frac{1}{2}\, [W,\Lambda_{\rm N}] + \frac{1}{8} \, [W,[W,\Lambda_{\rm N}]]$. Expanding the total action to the quadratic order of $W$ and requiring $\{W,\Lambda_{\rm N}\}=0$, $S_{\mathrm{tot}}[\Delta,Q]$ becomes $S_{\mathrm{tot}}[\Delta,\theta,W]$, where
\begin{align}
S_{\mathrm{tot}}[\Delta,\theta,W] &= S_0[W] + S_{\rm MF}[\theta,\Delta] + \tilde{S}_{\rm int}[\theta,W] \notag \\
&\quad + S_{\rm I}[\theta,W] + S_{\rm II}[\theta,\Delta,W]. \label{S_total_r}
\end{align}
Here, $S_0[W]$ is the action of the free cooperon and diffuson modes. We will cover the details of deriving these modes in the next section. The explicit expressions for each term except the last term in Eq.~(\ref{S_total_r}) are listed in Appendix \hyperref[AppendixB]{B}. The last term $S_{\rm II}$ contains terms in the form $\sim \mathrm{Tr}\,\{(\dots)WW\}$ and $\sim \mathrm{Tr}\,\{W(\dots)W\}$, which do not make any contribution in the first order of $\Delta_{\rm q}$ and thus are neglected.

Now, we can find the effective action of the gap function, $S_{\rm eff}$. Regarding the terms other than $S_0[W]$ as perturbative terms, $S_{\rm eff}$ is given as $\exp \lr{iS_{\rm eff}} = \braket{\exp\lr{i\lra{ S_{\rm MF} + S_{\rm int} + S_{\rm I}}} }_{0}$, where $\braket{\cdots}_0$ denotes the average with respect to $S_{0}$. Performing the cumulant expansion at the one-loop level, the effective action is obtained from the following equation:
\begin{equation}
S_{\rm eff } = \braket{S_{\rm MF} + \tilde{S}_{\rm int}}_{0} + \frac{i}{2}\braket{S^2_{\rm I}}_{0}. \label{Seff}
\end{equation}

\subsection{Parametrization of diffusive modes}
\label{MethodB}

In this section, we describe the parameterization of the field $W$ to calculate the average of the action with respect to $S_0$ in Eq. (\ref{Seff}). Note that the constraint on the field $W$ mentioned in the previous section, namely 
\begin{equation}
\{W,\Lambda_{\rm N}\}=0, \label{Wci}
\end{equation} 
is a sufficient condition to prevent formal divergence when integrating $\exp\,\{iS_0\}$ with respect to $W$ \cite{Finkelstein, Kamenev1, Feigelman}. The field $W$ can be parameterized as follows \cite{Feigelman}:
\begin{equation}
W=\begin{pmatrix}
\check{C}_{\rm R} & \check{B} \\
\check{\bar{B}} & \check{C}_{\rm A}  
\end{pmatrix}. \label{W}
\end{equation}
Here, $\check{C}_{\rm R}$ and $\check{C}_{\rm A}$ are responsible for the retarded and advanced cooperon modes, respectively, and $\check{B}$ and $\check{\bar{B}}$ are the normal and conjugated diffuson modes, respectively. Unlike the original work \cite{Feigelman}, here we consider the spin space to take into account the Rashba SOC and the Zeeman field. From Eqs.~(\ref{Wci}) and (\ref{W}), we obtain the retarded cooperon mode as $\check{C}_{\rm R} = \tau_- \hat{C}_{\rm R} - \tau_+ \hat{C}^{\dagger}_{\rm R}$ with
\begin{align}
\hat{C}_{\rm R} &= ic^{\rm R}_{0} -c^{\rm R}_{1}\sigma_1 + ic^{\rm R}_{2}\sigma_2 + ic^{\rm R}_{3}\sigma_3,  \label{CR}
\end{align}
where $c^{\mathrm{R}}_i$ are complex fields. Since the field $c^{\rm R}_i$ is expanded over all Pauli matrices, the above equation includes both the singlet and triplet channels of cooperon modes. The advanced cooperon mode $\check{C}_{\rm A}$ is obtained in the same manner. The normal diffuson mode is given as $\check{B} = \mathrm{diag}\,\{ \hat{B}^{\rm u},\hat{B}^{\rm d} \}$, where
\begin{align}  
\hat{B}^{\rm u(d)}&= \pm b^{\rm u(d)}_{0} - ib^{\rm u(d)}_{1}\sigma_1 + b^{\rm u(d)}_{2}\sigma_2 + b^{\rm u(d)}_{3}\sigma_3, \label{BB}
\end{align}
and $b^{\rm u(d)}_{i}$ are complex fields. The conjugated diffuson mode is obtained from the relation $\check{\bar{B}}=-\check{B}^{\dagger}$. As the same as the cooperon modes, the above equation includes both the singlet and triplet diffuson modes. Substituting these parameterizations into the bare effective action $S_0$, namely Eq.~(\ref{S0}), the equation can be seperated as
\begin{align}
S_0[W] &= \lrb{S_0[W]}_{\mathrm{Cooperon}}^{\rm R} + \lrb{S_0[W]}_{\mathrm{Cooperon}}^{\rm A} \notag \\
&\quad+ \lrb{S_0[W]}_{\mathrm{Diffuson}}^{\rm U} + \lrb{S_0[W]}_{\mathrm{Diffuson}}^{\rm D},
\end{align}
where
\begin{widetext}

\begin{align}
\lrb{S_0[W]}_{\mathrm{Cooperon}}^{\rm R(A)} &= \frac{i\pi\nu}{2} \sum_{\eps,\eps'} \Phi^{\rm cR(A) \dagger}_{\eps\eps'}(-\bm{q}) \,  \mathcal{D}^{-1}_{\rm cR(A)}(\tilde{\bm{q}},\eps+\eps') \, \Phi^{\rm cR(A)}_{\eps'\eps} (\bm{q})   ,         \label{Seff_Cooperon} \\
\lrb{S_0[W]}_{\mathrm{Diffuson}}^{\rm U(D)} &= \frac{i\pi\nu}{2} \sum_{\eps,\eps'} \Phi^{\rm dU(D)}_{\eps\eps'}(-\bm{q}) \,  \mathcal{D}^{-1}_{\rm d}(\bm{q},\eps-\eps') \, \bar{\Phi}^{\rm dU(D)}_{\eps'\eps} (\bm{q}), \\
\Phi^{\rm cR(A)}_{\eps\eps'}(\bm{q}) &= (c^{\rm R(A)}_{0,\eps\eps'}(\bm{q}), c^{\rm R(A)}_{1,\eps\eps'}(\bm{q}),   c^{\rm R(A)}_{2,\eps\eps'}(\bm{q}),  c^{\rm R(A)}_{3,\eps\eps'}(\bm{q})  )^{\rm t}, \\
\Phi^{\rm dU(D)}_{\eps\eps'}(\bm{q}) &= (b^{\rm U(D)}_{0,\eps\eps'}(\bm{q}), b^{\rm U(D)}_{1,\eps\eps'}(\bm{q}),   b^{\rm U(D)}_{2,\eps\eps'}(\bm{q}),  b^{\rm U(D)}_{3,\eps\eps'}(\bm{q})  )^{\rm t},
\end{align}
\end{widetext}
and $\tilde{q}_i=q_i+2A_i$. Here, the $4\times4$ matrices $\mathcal{D}^{-1}_{\rm cR(A)}$ and $\mathcal{D}^{-1}_{\rm d}$ are the inverses of the retarded (advanced) cooperon and diffuson propagators, respectively. Due to the length of their expressions, we show the explicit forms of $\mathcal{D}^{-1}_{\rm cR(A)}$ and $\mathcal{D}^{-1}_{\rm d}$ in Appendix \hyperref[AppendixC]{C}. Note that, unlike free cooperon modes, free diffuson modes are unaffected by the \textit{ordinary} gauge fields. This feature is shared with the previous result \cite{Kamenevhon}, which neglects the SOC.

\subsection{SC saddle point} 
\label{MethodC}

Choosing the appropriate saddle point is key to a good approximation of perturbation theory. It is naively expected that the best SC saddle point $\bar{Q}_{\rm s}$ should optimize the effective action Eq.~(\ref{Seff}). However, an estimation of $\bar{Q}_{\rm s}$ is necessary to make the analysis possible. This is achieved by setting a saddle point and then incorporating contributions from terms unrelated to the saddle point. We choose a saddle point that includes the Zeeman coupling. In the ordinary MF theory, this is obtained by solving equation $[ \check{E}_{\eps} + \check{\Delta}_{\rm cl} + \gamma_{\rm q} \check{\Delta}_{\rm q}, \bar{Q}_{\mathrm{s},\eps\eps}] = 0$ with $\Delta_{\rm q}=0$, where $\check{E}_{\eps}=\tau_3(\eps+h\sigma_3)$. However, the solution does not include any effect of the terms unrelated to the saddle point because the quantum part necessary for deriving the gap equation is absent. To overcome this, we demand the retarded and advanced components of the equation to be zero, i.e.,
\begin{align}
&\quad\quad~[ \check{E}_{\eps} +  \check{\Delta}_{\mathrm{R(A)},\eps}, \check{\Lambda}_{\mathrm{sR(A)},\eps\eps}] = 0,   \label{Saddle2}     \\           
& \check{\Delta}_{\mathrm{R(A)},\eps}=i\tau_2\Delta_{\mathrm{R(A)},\eps} \equiv i\tau_2\,(\Delta_{\rm cl}  \pm F_{\eps}\Delta_{\rm q}). \label{Delta}
\end{align}
Solving Eq.~(\ref{Saddle2}), we obtain the saddle point $\bar{Q}_{\rm s} = U_{\rm K}\Lambda_{\rm s} U_{\rm K}$ with $\Lambda_{\rm s} = \mathrm{diag}\,\{\check{\Lambda}_{\rm sR},\check{\Lambda}_{\rm sA}\}$, which can be written in the form $\check{\Lambda}_{\rm sR(A)} = \check{U}_{\rm sR(A)} \tau_3 \check{U}^{-1}_{\rm sR(A)}$. By some calculations, we obtain $\check{U}_{\rm sR(A)} = 1 - \tau_1 \hat{X}_{\rm R(A)}$, where
 \begin{align}
\hat{X}_{\rm R} =  \mathrm{diag}\,& \Big\{ \tanh \frac{\theta_{\mathrm{R},\eps+}}{2}, \tanh \frac{\theta_{\mathrm{R},\eps-}}{2} \Big\}, \label{spectralUX} \\
\tanh  \theta_{\mathrm{R},\eps\pm} &= \mathrm{sgn}(\eps_{\pm})\Delta_{\mathrm{R},\eps}/|\eps_{\pm} \!+ \! i0|, \label{spectral1}
\end{align}
and $\eps_{\pm}=\eps\pm h$. $\hat{X}_{\rm A}$ has the same form as $\hat{X}_{\rm R}$ but $\theta_{\mathrm{R},\eps\pm}$ is replaced by $\theta_{\mathrm{A},\eps\pm}$, which satisfies $\tanh\theta_{\mathrm{A},\eps\pm}= \Delta_{\mathrm{A},\eps}/|\eps_{\pm} \!-\! i0|$. This is indeed a possible form of the \textit{spectral angle} mentioned below Eq.~(\ref{Ukeys}). Defining $U\equiv U_{\rm K}U_{\rm s}$ and $U_{\rm s} \equiv \mathrm{diag}\lrb{\check{U}_{\rm sR},\check{U}_{\rm sA}}$, we can rewrite $\bar{Q}_{\rm s}$ as $\bar{Q}_{\rm s} = U\Lambda_{\rm N} U^{-1}$.

The solution presented above can be used to obtain the effective action incorporating the contributions of the SOC, Cooper-pair momentum, and interactions. This is allowed by the presence of $F_{\eps}$ in Eq.~(\ref{Delta}). Without this term, the effective action would become trivial through the energy integration. However, since we search the saddle point that optimizes Eq.~(\ref{Seff}), some terms in Eq.~(\ref{spectralUX}) need to be specified by the saddle point condition for Eq.~(\ref{Seff}). Similarly to Ref.~\cite{Burmistrov2}, we leave the spectral angle as an unknown quality. By paying attention to the role of $F_{\eps}$ described above, we arrive at the ansatz that the spectral angles in Eq.~(\ref{spectralUX}) should be replaced by the \textit{generalized} spectral angles shown below:
\begin{align}
\theta_{\mathrm{R},\eps\pm} \! &\rightarrow \theta_{\mathrm{R},\eps\pm} = \theta^{\rm cl}_{\mathrm{R},\eps \pm }\!\!+F_{\eps}\theta^{\rm q}_{\mathrm{R},\eps \pm},        \label{spectralR} \\
\theta_{\mathrm{A},\eps\pm} \! &\rightarrow \theta_{\mathrm{A},\eps\pm} = \theta^{\rm cl}_{\mathrm{A},\eps \pm }\!\!-F_{\eps}\theta^{\rm q}_{\mathrm{A},\eps \pm},  \label{spectralA}
\end{align}
where $\theta^{\rm cl}_{\mathrm{R(A)},\eps \pm}$ and $\theta^{\rm q}_{\mathrm{R(A)},\eps \pm}$ represent the classical and quantum parts of the generalized spectral angles, respectively. Hereafter, we adopt the SC saddle point $\bar{Q}_{\rm s}$ that is determined by Eq.~(\ref{spectralUX}) with Eqs.~(\ref{spectralR}) and (\ref{spectralA}). The quantum parts will be set to zero after performing the first derivatives with respect to them.

It is instructive to point out that the retareded and advanced parts of the SC saddle point $\bar{Q}_{\rm s}$, which are $\check{\Lambda}_{\rm sR}$ and $\check{\Lambda}_{\rm sA}$, respectively, can be split into singlet and triplet sectors. It is sufficient to show only the retarded part. Using Eq. (\ref{spectralUX}), we arrive at the expression $\check{\Lambda}_{\rm sR}=i\tau_2\hat{\Lambda}_{\rm 2R} + \tau_3\hat{\Lambda}_{\rm 3R}$, where
\begin{align}
\hat{\Lambda}_{\rm 2R} &= \mathrm{diag}\,\{ \sinh\theta_{\rm R,\eps+}, \sinh\theta_{\rm R,\eps-}       \}, \\
\hat{\Lambda}_{\rm 3R} &= \mathrm{diag}\,\{ \cosh\theta_{\rm R,\eps+}, \cosh\theta_{\rm R,\eps-}       \}.
\end{align}
Note that, prior to reaching $\bar{Q}_{\rm s}$, the rotation transformation with respect to the matrix $\check{L}_{\rm r}$ [see below Eq. (\ref{Sp})] and the U(1) transformation [see Appendix \hyperref[AppendixA]{A}] had been performed on each matrix in Eq. (\ref{SQ_tot}). Nevertheless, to achieve our purpose here, it is sufficient to carry out only the inverse transformation with respect to $\check{L}_{\rm r}$. Let $\check{\Lambda}'_{\rm sR}\equiv \check{L}_{\rm r} \check{\Lambda}_{\rm sR}\check{L}_{\rm r}$. Then, we can show that
\begin{equation}
\check{\Lambda}'_{\rm sR} = \hat{g}_{\rm R,0} + \hat{g}_{\rm R,t} \sigma_1, \label{singtrip}
\end{equation}
where $\hat{g}_{\rm R, 0}$ and $\hat{g}_{\rm R,t}$ are singlet and triplet parts, respectively. They can be expressed as
\begin{align}
\hat{g}_{\rm R,0} &= \frac{1}{2}\sum_{\lambda=\pm}(\tau_3\cosh\theta_{\rm R,\eps\lambda} + i\tau_2\sinh\theta_{\rm R,\eps\lambda} ), \\
\hat{g}_{\rm R,t} &= \frac{1}{2}\sum_{\lambda=\pm}\lambda(\tau_3\cosh\theta_{\rm R,\eps\lambda} + i\tau_2\sinh\theta_{ \rm R,\eps\lambda} ).
\end{align}
Note that the form of Eq. (\ref{singtrip}) is the same as that in Ref. \cite{Ilic2}, except that the Pauli matrix of the triplet part is $\sigma_1$ in our study because the magnetic field is in the $x$-direction. Furthermore, the triplet sector with the value of $\theta_{\rm R(A),\eps\pm}$ calculated in the next section does not vanish in general at finite magnetic fields, even when the WL correction is not considered. 

\subsection{Modified Usadel equation} 
\label{MethodD}

Following the plan described in the previous paragraph, we determine the form of $\theta^{\rm c}_{\mathrm{R(A)},\eps\pm}$ by optimizing the effective action $S_{\rm eff}$ with respect to each spectral angle. We work within the GL theory by expanding $\theta_{\mathrm{R(A)},\eps\pm}$ as $\theta^{\rm cl}_{\mathrm{R(A)},\eps\pm} = \Delta_{\rm cl} f^{\rm R(A)}_{1\eps\pm} + \Delta^3_{\rm cl} f^{\rm R(A)}_{3\eps\pm}$ and $\theta^{\rm q}_{\mathrm{R(A)},\eps\pm} = \theta^{\rm R(A)q}_{1\eps\pm} + \theta^{\rm R(A)q}_{3\eps\pm}$. In this way, $f^{\rm R(A)}_{j\eps\pm}$ is determined by the following equation:
\begin{equation}
[\delta S_{\rm{eff}}/\delta \theta^{\mu \mathrm{q}}_{j\eps\pm}]_{\theta^{\mathrm{q}}\rightarrow 0} = 0, \label{UsadelS}
\end{equation}
where $\mu =\rm R,A$ and $j=1$ or $3$. In the following, $f^{\rm R(A)}_{j\eps\pm}$ is referred to as \textit{spectral function}. Note that this spectral function $f^{\rm R(A)}_{j\eps\pm}$ is different from the Fermi distribution function. The retarded and advanced spectral functions are not actually independent. Otherwise, irregular divergences appear in the left-hand side of Eq.~(\ref{UsadelS}). This argument is explained in detail in the top section of Appendix \hyperref[AppendixD]{D}. We find that all irregular terms vanish if the following condition holds:
\begin{align}
f^{\rm R}_{j\eps\pm} = -f^{\rm A}_{j-\eps\mp},\quad \theta^{\rm Rq}_{j\eps\pm} = -\theta^{\rm Aq}_{j-\eps\mp}.  \label{spectralci}
\end{align}
Using Eqs.~(\ref{spectralR})$-$(\ref{spectralci}) and performing analytic continuation, we obtain the \textit{modified Usadel equation} \cite{Burmistrov2} in the first and third orders of $\Delta_{\rm cl}$ as follows:
\begin{equation}
\mathcal{S}_{\mathrm{kernel},j}[f_{1n\pm},f_{3n\pm}] = \mathcal{S}_{\mathrm{source}, j}[f_{1n\pm},f_{3n\pm}], \label{Modified_Usadel}
\end{equation}
where $f_{jn\pm} = if^{\rm R}_{ji\omega_n\pm} $, $\omega_{n\pm} = \omega_n\pm ih$, $\omega_n = (2n+1) \pi T~(n\in \mathbb{Z})$, and $\mathcal{S}_{\mathrm{source},1}=1$. The others are shown in Appendix \hyperref[AppendixD]{D} as Eqs. (\ref{Usadel1d}), (\ref{Usadel3dk}), and (\ref{soc3}) to avoid describing lengthy equations. The details of the calculation are described there as well. We find that the correction from the long-range Coulomb interaction is much greater than that from the Cooper interaction. The equations are solved numerically by the Pad\'{e} decomposition method~\cite{Ozaki}, which can achieve fast convergence when calculating the Matsubara summation.

 %\YYS{The momentum integrations in the modified Usadel equations are bounded by the mean-free path $l_F=v_F\tau$.} 

%where $f_{a, n\pm} = if^R_{a,i\omega_n\pm} $, $a = 1,3$, $\omega_{n\pm} = \omega_n\pm ih$, $\omega_n = (2n+1) \pi T~(n\in \mathbb{N})$, and $\beta=\tau/m$. 

%$\Gamma_s/(4\pi) = \pi^2\nu\tilde{\Gamma}_s/16,~[\Gamma_s/(2\pi)][2/(\pi\nu)^2] = \tilde{\Gamma}_s/(4\nu)$, and $\tilde{\Gamma}_s=1$.
%$C_0$ is an unimportant constant determined by comparing our result for the case of conventional superconductors to the result in Ref. \cite{Levchenko1}. 

\subsection{Equation of state} 
\label{MethodE}

The gap equation is obtained from the stationary condition $[\partial S/\partial \Delta_{\rm q}]_{\Delta_{\rm q} \rightarrow 0}=0$. At the GL level, the equation is in the form $\alpha_2\Delta_{\rm cl} + \alpha_4\Delta_{\rm cl}^3=0$, which corresponds to minimizing the GL free energy $f_{\mathrm{GL}} = \mathcal{C}\,(\frac{1}{2} \alpha_2\Delta^2_{\rm cl} + \frac{1}{4}\alpha_4\Delta^4_{\rm cl})$, where $\mathcal{C}$ is a numerical constant. The coefficients $\alpha_2$ and $\alpha_4$ are given as follows:
\begin{align}
\alpha_2 &= 4\, \Gamma^{-1}_{\rm c} \! - 4\pi T \! \sum_{n>0}\sum_{\lambda=\pm}\Big[ f_{1n \lambda} - \frac{1}{\omega_n \!+\! i\lambda h}  \Big] \notag \\ 
&\quad - 2\int_{0}^{\omega_{\rm D}} \frac{d\eps}{\eps} \Big[ \tanh \lr{\frac{\eps_+}{2T}}   +     \tanh \lr{\frac{\eps_-}{2T}}  \Big], \label{alpha2} \\
\alpha_4&= -4\pi T \! \sum_{n>0}\sum_{\lambda=\pm} \Big( f_{3n \lambda} -\! \frac{1}{6}[f_{1n \lambda}]^3 \Big), \label{alpha4}
\end{align}
where $\Gamma_{\rm c}$ is the SC coupling constant. Note that $f_{jn\lambda}$ is derived from Eq.~(\ref{Modified_Usadel}), and $\alpha_2$ is regularized by introducing the ultraviolet cutoff, which is the Debye energy $\omega_{\rm D}$, to the energy integration. Since we are considering helical SC states, the supercurrent $I_s$ in the $y$-direction is given as follows \cite{Daido1}:
\begin{equation} 
I_{\rm s}=\partial f_{\mathrm{GL}}/\partial A_y= -\frac{1}{4}\mathcal{C}\frac{\partial }{\partial A_y}\,(\alpha^2_2/\alpha_4).
\end{equation} 
The second-order SC phase transition line in the absence of the supercurrent is calculated based on the conditions $\alpha_2 = 0$ and $\partial\alpha_2/\partial A_y =0$. We denote the transition temperature at a magnetic field $h$ as $T_{\rm c}(h)$, which is denoted as $T_{\rm c0}(h)$ when we neglect the e-e interaction. For simplicity we write $T_{\rm c0}(0)=T_{\rm c0}$ and $T_{\rm c}(0) = T_{\rm c}$. Furthermore, we calculate the critical currents $I_{\rm sc\pm}$ in the $\pm y$-direction, which is obtained by optimizing $I_{\rm s}$, with focus on the diode quality factor defined as $\eta = (1 -\xi)/(1+\xi)$ with $\xi = |I_{\rm sc-}/I_{\rm sc+}|$. 

\section{Numerical Results}\label{Numerical}

In the numerical calculations, we fix the parameters, $\Gamma_{\rm c} = 0.2$, $\omega_{\rm D}/E_{\rm F} = 0.1309$, and Fermi energy $E_{\rm F}/T_{\rm c0} = p_{\rm F}^2/(2mT_{\rm c0})= 10^3$, where $p_{\rm F}=mv_{\rm F}$ and $m$ is the electron mass. For the Rashba SOC constant $\alpha$ and impurity scattering time $\tau$, we consider the following two sets of parameters: (I) $\alpha/p_{\rm F} = 0.01,~\tau T_{\rm c0} = 1.5\times10^{-2}$ and (II) $\alpha/p_{\rm F} = 0.002,~\tau T_{\rm c0} = 7.5\times10^{-3}$. Note that the characteristic energies of SOC and disorder are represented by $\eps_{\rm SOC}=\alpha v_{\rm F}$ and $\eps_{\rm D}=1/\tau$, respectively. Therefore, case (I) is considered a comparable SOC case with $\eps_{\rm SOC}/\eps_{\rm D}=0.3$, while case (II) is considered a weak SOC case with $\eps_{\rm SOC}/\eps_{\rm D}=0.03$. Note also that our perturbative approach is valid for large dimensionless conductance $g \gg 1$ \cite{Andriyakhina1, Burmistrov2}. Here, $g$ is given as $g=2\pi \sigma_{\rm D}$ \cite{Andriyakhina2}, where $\sigma_{\rm D}=2\nu D$ is the Drude conductivity and $\nu = m/(2\pi)$ is the 2D density of states \textit{per spin}. We have confirmed that the impurity scattering time $\tau$ in both parameter sets (I) and (II) satisfies this condition.

%We adopt the MF theory for spatially uniform $s$-wave states.

%In what follows, we refer to the case that includes the corrections due to disorder and interactions as the interaction case, and to the case that excludes them as the free case.

%Since there is no contribution from the Cooper channel in the equation determining the second-order SC transition line, the suppression of the transition temperature is attributed to the Coulomb channel. 

%In Fig. \hyperref[fig1]{1a}, $(T^*_{\rm c0},h^*_{\rm c0}) \simeq (0.14T_{\rm c0}, 2.6T_{\rm c0})$ in the free case and  $(T^*_{\rm c},h^*_{\rm c}) \simeq (0.09T_{\rm c0}, 2.4T_{\rm c0})$. In Fig. \hyperref[fig3]{3a}, $(T^*_{\rm c0},h^*_{\rm c0}) \simeq (0.55T_{\rm c0}, 1.1T_{\rm c0})$ in the free case and  $(T^*_{\rm c},h^*_{\rm c}) \simeq (0.32T_{\rm c0}, 0.7T_{\rm c0})$. 

\subsection{Phase diagram}\label{NumericalA}

\begin{figure}[t]

\centering

\includegraphics[width=8.7cm]{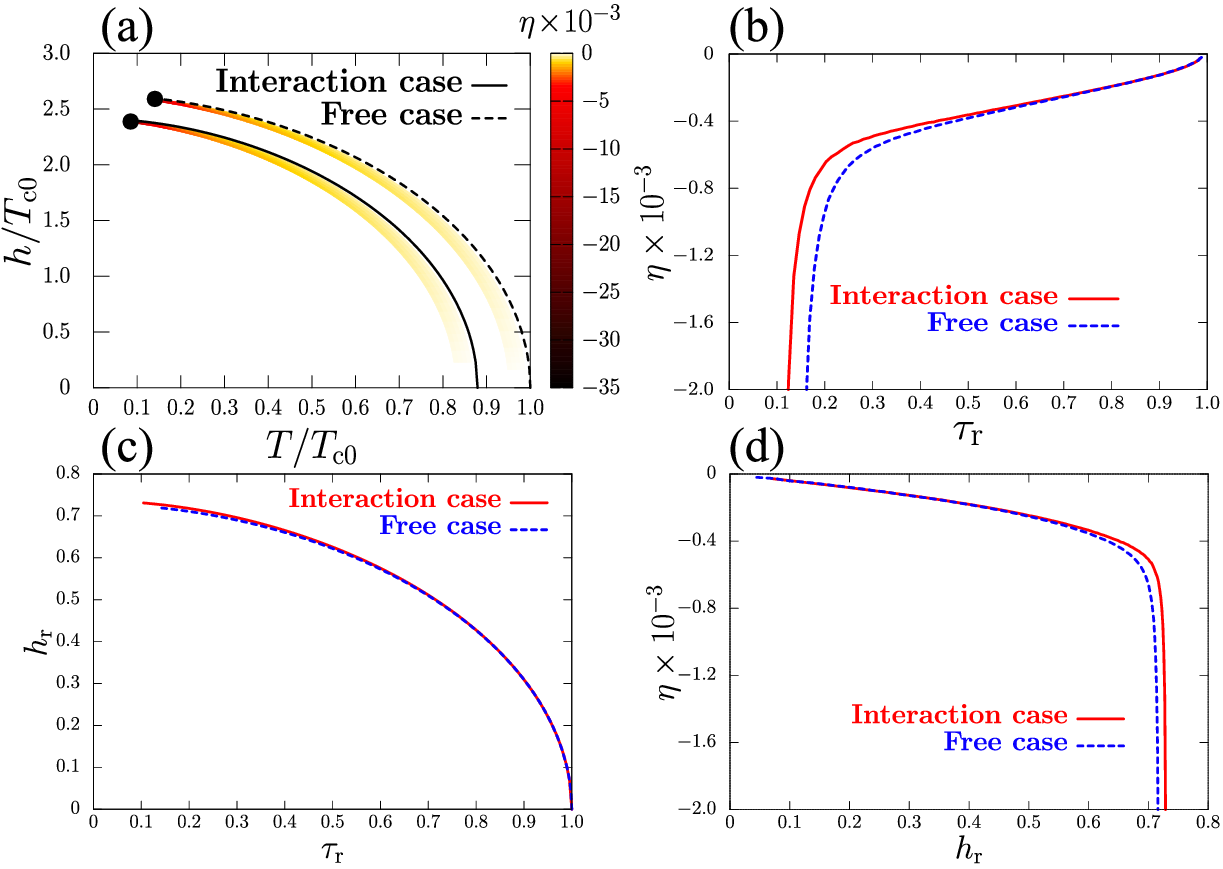}

\caption{The phase diagram and diode quality factor $\eta$ in the comparable SOC case ($\alpha/p_{\rm F} = 0.01,~\tau T_{\rm c0} = 1.5\times10^{-2}$). The solid (dashed) lines show the results in the case with (without) e-e interactions, which is referred to as the interaction (free) case in the main text. (a) The color maps of $\eta$ in the vicinity of the second-order SC transition lines, which change to first-order transitions at the tricritical points represented by the black circles. (c) The transition lines from panel (a) on the normalized temperature-normalized magnetic field scale, namely the $\tau_{\rm r}$-$h_{\rm r}$ scale. (b) and (d) the diode quality factor $\eta$ as a function of $\tau_{\rm r}$ and $h_{\rm r}$, respectively. In these figures, we vary $\tau_{\rm r}$ and $h_{\rm r}$ so that the temperatures satisfy $T=0.99T_{\rm c0}(h)$ in the free case and $T=0.99T_{\rm c}(h)$ in the interaction case. Note that the same $\tau_{\rm r}$ does not imply the same $h$ or $h_{\rm r}$ between the interaction and free cases.}

\label{fig1}

\end{figure}

We first show the properties of zero-current SC states. As shown in Figs.~\hyperref[fig1]{1a} and \hyperref[fig3]{3a}, the transition temperatures and magnetic fields are lower in the interaction case than in the free case. The gauge field of the helical SC state at zero current is also suppressed when the interactions are incorporated~[see Appendix \hyperref[AppendixE]{E}]. The zero-field transition temperatures are $T_{\rm c}=0.8793T_{\rm c0}$ and $T_{\rm c}=0.6457T_{\rm c0}$ in Figs.~\hyperref[fig1]{1a} and \hyperref[fig3]{3a}, respectively. We have confirmed that $T_{\rm c}$ decreases as the impurity scattering time $\tau$ decreases, while it increases toward $T_{\rm c0}$ as the SOC $\alpha$ increases [see Appendix \hyperref[AppendixF]{F}]. The second-order transition lines become first-order transitions at the tricritical points, where $\alpha_4$ becomes negative. The tricritical temperature and magnetic field in the interaction case, $T^*_{\rm c}$ and $h^*_{\rm c}$, are lower than those in the free case, $T^*_{\rm c0}$ and $h^*_{\rm c0}$, respectively. This first-order transition is well known to occur in the high field region due to the Pauli paramagnetic effects~\cite{Maki}, and it also occurs when the SOC is sufficiently weak, namely $\Gamma_{\rm r}= 2D\alpha^2 \lesssim 10 \Delta_0$ where $\Delta_0\sim 1.8T_{\rm c0}$ is the SC gap at zero temperature~\cite{Ilic2, Heikkila}. The parameters used here satisfy this condition. At magnetic fields higher than the tricritical points, although a second-order transition to a stripe SC state is expected in the clean limit~\cite{Agterberg}, it turns out to be a first-order transition to a uniform SC state in the dirty limit~\cite{Buzdin}. Unlike the previous study \cite{Ilic2} that solves the MF gap equation rigorously, our approach cannot describe the first-order transition because the $\Delta_{\rm cl}$-term in the free energy is up to the 4th order.

\subsection{Superconducting diode effect}\label{NumericalB}

Let us discuss the SD effect with focus on the diode quality factor $\eta$. Since $\eta$ is tiny in the weak SOC case, we calculate the SD effect in the comparable SOC case and show the color map of $\eta$ in Fig.~\hyperref[fig1]{1a}. It appears that $\eta$ increases as the system approaches the tricritical point. To show the behavior of $\eta$ near the transition line, we plot in Fig.~\hyperref[fig1]{1b} the diode quality factor near the transition line as a function of normalized temperature $\tau_{\rm r}$, defined as $\tau_{\rm r} = T/T_{\rm c0}$ or $\tau_{\rm r} = T/T_{\rm c}$. We see qualitatively the same behavior of $\eta$ in the absence and presence of the e-e interaction. Especially, $\eta$ in the two cases almost coincide at high $\tau_{\rm r}$ despite the significant decrease in the zero-field transition temperature due to the e-e interactions. This means that the SD effect with the same order of $\eta$ as in the free case can be achieved in the interaction case. This finding also suggests a universal law of the diode quality factor, which will be derived below.

\captionsetup{font=normal,justification=raggedright,singlelinecheck=false}

 \begin{figure}[t]

\centering

\includegraphics[width=8.7cm]{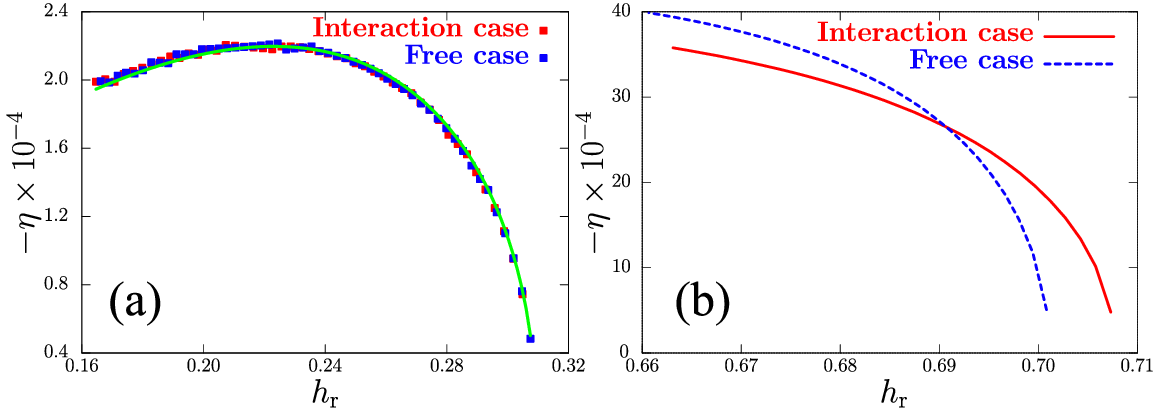}

\caption{The diode quality $\eta$ is plotted as a function of the normalized magnetic field $h_{\rm r}$ at normalized temperatures: (a) $\tau_{\rm r}=0.9$ and (b) $\tau_{\rm r}=0.25$. In panel (a), the red and blue squares denote $\eta$ in the interaction and free cases, respectively. The green line shows the fitting by Eq.~(\ref{Diode}). In panel (b), the red solid and blue dashed lines show $\eta$ in the interaction and free cases, respectively.}

\label{fig2}

\end{figure}

In the GL theory \cite{Yuan,Daido1}, the diode quality factor near the second-order transition line follows the scaling law $\eta \sim \mathcal{F}\lra{1-\tilde{\tau}}^{1/2}$. In the free case, the reduced temperature is $\tilde{\tau} = T/T_{\rm c0}(h)$, and the coefficient can be expanded by odd powers of the magnetic field, $\mathcal{F} = \sum_{n=0} A_{2n+1}h^{2n+1}$. In the interaction case, we replace $T_{\rm c0}(h)$ and $A_{2n+1}$ with $T_{\rm c}(h)$ and $A'_{2n+1}$, respectively. The relation between $A_1$ and $A'_1$ is derived from the following analysis. We can fit transition lines in Fig.~\hyperref[fig1]{1a} by $ T_{\rm c0}(h)/T_{\rm c0} = [1-\chi_{\rm c0} (h/h_{\rm c0})^2]^{1/\chi_{\rm c0}}$ and $ T_{\rm c}(h)/T_{\rm c} = [1-\chi_{\rm c} (h/h_{\rm c})^2]^{1/\chi_{\rm c}}$. These reduce to the known behavior of the transition line $\tau_{\rm r} \simeq 1- h^2_{\rm r}$ at low fields, where $h_{\rm r} \equiv h/h_{\rm c0}$~$(h/h_{\rm c})$ and $\tau_{\rm r} = T/T_{\rm c0}$~$(T/T_{\rm c})$. Thus, the transition lines almost coincide on the $\tau_{\rm r}$-$h_{\rm r}$ scale at low fields [Fig.~\hyperref[fig1]{1c}]. We find $h_{\rm c0}=3.61T_{\rm c0}$, $h_c=3.28T_{\rm c0}$, $\chi_{\rm c0} = 1.89$, and $\chi_c=1.84$. Note that $h_{\rm c0}$ and $h_{\rm c}$ are not equivalent to the tricritcal fields ($h^*_{\rm c0} = 2.6T_{\rm c0}$ and $h^*_{\rm c} = 2.4T_{\rm c0}$). When the horizontal axis in Fig.~\hyperref[fig1]{1b} is changed to $h_{\rm r}$, the diode quality factors nearly overlap at low fields, as shown in Fig.~\hyperref[fig1]{1d}. This means that $A_1/h_{\rm c0}=A'_1/h_{\rm c}\equiv \mathcal{A}_1$, and $\eta$ at low fields follows the universal behavior
\begin{equation}
\eta =  \mathcal{F} \big[1- \tilde{\tau} \big]^{1/2},\quad \mathcal{F} \simeq \mathcal{A}_1h_{\mathrm r}. \label{Diode}
\end{equation}
We further find that this law is valid in a wide range of the phase diagram at high temperatures $\tau_{\rm r}>0.9$, as demonstrated in Fig.~\hyperref[fig2]{2a} by setting $\mathcal{A} = 2.8\times 10^{-3}\tau^{-1}_{\rm r}$. Based on these analyses, we conclude that the diode quality factor is robust against the corrections due to disorder and e-e interactions in the low-magnetic-field regime. 

In previous work \cite{Hasan}, a system essentially similar to ours was studied without incorporating e-e interactions. The linear field dependence of $\eta$ is found to be due to the singlet-triplet mixing mechanism. However, unlike our Eq.~(\ref{Diode}), the analytical expression derived in Ref.~\cite{Hasan} cannot describe the whole behavior of $\eta$ in Fig.~\hyperref[fig2]{2a}. Nevertheless, combining their results and Eq.~(\ref{Diode}) suggests that the interactions may lower both Rashba SOC and transition temperature while preserving the SD effect.  

In the low-temperature regime, deviation from the universal law manifests, as shown in Fig.~\hyperref[fig2]{2b}. The roles of the e-e interactions appear non-monotonically. In the high-magnetic-field regime, the diode quality factor $\eta$ is higher in the interaction case than in the free case. This is a result of an increase in the tricritical field in the $h_{\rm r}$ scale due to e-e interactions, and can be qualitatively explained by Eq.~(\ref{Diode}). However, at low magnetic fields, the diode quality factor is decreased by the interaction.

\subsection{WL conductivity}\label{NumericalC}

When the supercurrent reaches its critical value, the SC state changes to a resistive state. Here, we study the quantum corrections to conductivity in the resistive states. For simplicity, we neglect the corrections to conductivity due to e-e interactions and consider the conventional WL conductivity $\sigma_{\rm WL}$~\cite{Hikami2, Maekawa2, Bergmann}. The calculation of $\sigma_{\rm WL}$ can be performed in the Keldysh functional formalism by extending the previous work \cite{Kamenevhon} that considers only the spinless case to include the spin degrees of freedom~[see Appendix \hyperref[AppendixG]{G}]. It appears that our equation for $\sigma_{\rm WL}$ is consistent with existing results \cite{Kashuba, Marinescu, ilic2, Burmistrov3}.

 \begin{figure}[t]

\centering

\includegraphics[width=8.7cm]{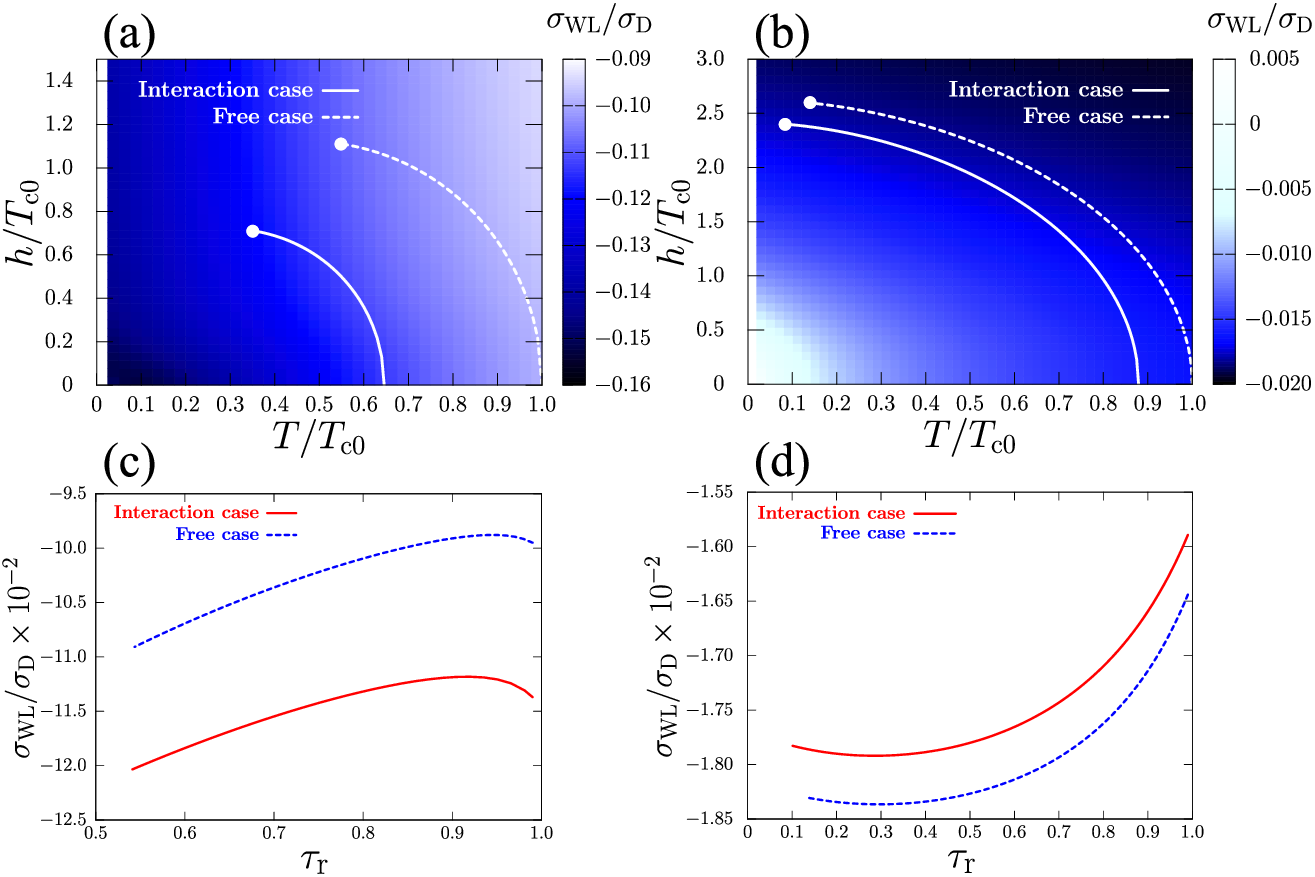}

\caption{The WL conductivity $\sigma_{\rm WL}$ in (a) the weak SOC regime ($\alpha/p_{\rm F} = 0.002,~\tau T_{\rm c0} = 7.5\times10^{-3}$) and (b) the comparable SOC case ($\alpha/p_{\rm F} = 0.01,~\tau T_{\rm c0} = 1.5\times10^{-2}$). The solid (dashed) lines show the results in the interaction (free) case. The white circles represent the tricritical points. Panels (c) and (d) show $\sigma_{\rm WL}$ almost along the SC transition lines in panels (a) and (b), respectively. The data is plotted in the same way as in Fig. \hyperref[fig1]{1b}. The red solid (blue dashed) curves represent the interaction (free) cases. }

\label{fig3}

\end{figure} 

The numerical results of the WL conductivity $\sigma_{\rm WL}$ are shown in Fig.~\hyperref[fig3]{3}. The WL correction normalized by the Drude conductivity is much larger in the weak SOC case [Fig.~\hyperref[fig3]{3a}] than in the comparable SOC case [Fig.~\hyperref[fig3]{3b}]. As the temperature and magnetic field increase, the magnitude of $\sigma_{\rm WL}$ decreases in the weak SOC case but increases in the comparable SOC case. Note that the magnitude of $\sigma_{\rm WL}$ in the comparable SOC case does not diverge at extremely-high temperatures and magnetic fields [see Fig. \hyperref[fig3s]{6} in Appendix \hyperref[AppendixG]{G}]. Nevertheless, these results suggest that the WL behaviors are suppressed by the SOC irrespective of Coulomb interactions. In contrast, the magnitude of the diode quality factor $\eta$ with a large SOC is usually larger than with a small SOC for a given temperature and magnetic field. Therefore, our results reveal a trade-off relationship between the SD effect and the WL behavior in the resistive states, at least in the parameter range considered here.

The magnitude relation of the WL conductivity between the interaction and free cases is also different because superconductivity is suppressed by the long-range Coulomb interaction and the dependencies of $\sigma_{\rm WL}$ on temperature and magnetic field are different for these SOC strength. In the weak SOC case, the magnitudes of $\sigma_{\rm WL}$ in the interaction case become greater than those in the free case [Fig. \hyperref[fig3]{3c}], while the opposite behavior appears in the comparable SOC case [Fig. \hyperref[fig3]{3d}]. Moreover, the conductivity curves in the comparable SOC case are convex, while those in the weak one are concave.

\section{Summary and Discussion} 
\label{Summary}

We have formulated the renormalized GL theory including the Cooper and long-range Coulomb interaction terms from the NLSM, based on the Keldysh functional formalism. Our formalism extends the previous one and enables us to study dirty 2D Rashba superconductors under in-plane Zeeman fields, which are typical platforms of the SD effect. We found the robustness of the SD effect with respect to the Cooper and Coulomb interactions near the low-transition magnetic fields, even though the SC states and first-order transition are suppressed by the long-range Coulomb interaction in particular. This means that the magnetic field required to produce a diode quality factor can be reduced by the Coulomb interaction. In combination with the previous result \cite{Hasan}, the robustness of the SD effect implies the suppression of the Rashba SOC effect by the long-range Coulomb interaction. Nevertheless, the robustness is broken near the tricritical point located in the high-magnetic field regime. We also demonstrated a trade-off relationship between the WL behavior in the resistive state, which emerges when the current is beyond the critical current, and the efficiency of the SD effect. 

%The relationship is irrelevant to the interaction. Nevertheless, due to different dependencies of the WL behavior in the resistive state on temperature and magnetic field for each SOC strength, the Coulomb interaction may take a role to suppress (enhance) the WL behavior in the vicinity of the SC transition line if the SOC strength is comparable (weak). 

Although the result of the trade-off relationship is contrary to the initial expectation, it is too early to rule out the possibility of the Anderson insulating transition occurring after superconductivity is destroyed by the critical current. Note that our calculation was performed within a narrow range of parameters and that only a few channels of the e-e interaction were considered. Nevertheless, our formulation lays the foundation for investigating the cooperation between the SD effect and the localization effect in the future. Furthermore, since the long-range Coulomb interaction is ubiquitous in the solid state systems \cite{Raffy1983, Graybeal1984, Chand2012, Noat2013, Banerjee2017}, we expect our results to help realize the SD effect in the present setup and pave the way for switching between superconducting, metallic, and insulating behavior via an electric current.

\section*{Acknowledgements}
We appreciate helpful discussions with Ryusuke Ikeda, Hideki Narita, Teruo Ono, and Ryuji Hakuno. This work was supported by JSPS KAKENHI (Grant Numbers JP22H01181, JP22H04933, JP23K17353, JP23K22452, JP24K21530, JP24H00007, JP25H01249).

\appendix*

%\def\thesection{\Alph{section}}

%\section{Effective action\label{AppendixA}}

\begin{widetext}

\section*{Appendix A: Notation and Derivation of Model's Action}
\label{AppendixA}

\setcounter{equation}{1}
\setcounter{section}{1}
\def\thesubsection{\arabic{subsection}}

\subsection{Explicit forms of the trace and tensor product}

We present the explicit forms of the trace and several matrices in Eqs.~(\ref{Snr}) and (\ref{Sp}). The trace in those equations is represented in the space and time, for example, as
\begin{align}
\mathrm{Tr}\,\big\{    \Omega Q       \big\} \equiv \int dtd\bm{r}\, \mathrm{tr}\, \big\{\Omega(\bm{r},t)Q_{tt}(\bm{r})\big\}, \quad \mathrm{Tr}\,\big\{ (\hat{\nabla}_iQ)^2      \big\} \equiv \int dt_1dt_2d\bm{r}\, \mathrm{tr} \, \big\{ [\hat{\nabla}_iQ_{t_1t_2}(\bm{r})] [\hat{\nabla}_iQ_{t_2t_1}(\bm{r})] \big\},
\end{align}
where the matrix $Q_{t_1t_2}(\bm{r})$ and $\Omega(\bm{r},t)$ are the real-space-time representations of $Q$ and $\Omega$, respectively. For the trace $\mathrm{tr}$, see its definition below Eq. (\ref{Cooper_int}). Note that $\hat{\nabla}_i$ only depends on the spatial coordinate $\bm{r}$ because we have assumed that the gauge fields are uniform in both space and time. Regarding the tensor product, we use the abbreviation $AB \equiv A\otimes B$ for given matrices $A$ and $B$ as in Sec. \ref{Model}. Thus, the expressions of $\gamma_{\rm cl}\hat{\nabla}_i = \gamma_{\rm cl} \otimes \hat{\nabla}_i$, $\gamma_{\rm cl}\hat{E} = \gamma_{\rm cl}\otimes\hat{E}$, $\gamma_{\rm cl(q)}\hat{\bm{\Delta}}_{\rm cl(q)}=\gamma_{\rm cl(q)}\otimes\hat{\bm{\Delta}}_{\rm cl (q)}$, and $\gamma_{\rm cl}\hat{F}_{xy}=\gamma_{\rm cl}\otimes\hat{F}_{xy}$, which are indeed the tensor products of $\gamma_{\rm cl}$, $\tau_i$, and $\sigma_i$, are $8\times8$ matrices, consistent with the dimension of $Q$. The matrix form of $\gamma_{\rm cl (q)}\tau_i\sigma_j$ is given as follows:
\begin{equation}
 \gamma_{\rm cl}\tau_i\sigma_j \equiv \gamma_{\rm cl}\otimes \tau_{i} \otimes \sigma_j = \begin{pmatrix}
\tau_i\sigma_j & 0 \\
0 & \tau_i\sigma_j
\end{pmatrix},\quad \gamma_{\rm q}\tau_i\sigma_j  \equiv \gamma_{\rm q}\otimes \tau_{i} \otimes \sigma_j = \begin{pmatrix}
0& \tau_i \sigma_j \\
\tau_i\sigma_j & 0
\end{pmatrix}.
\end{equation}
The explicit form of $\tau_i\sigma_j$ is given for several $i$ and $j$ as follows:
\begin{equation}
\tau_1\sigma_2 = \begin{pmatrix}
0 & \sigma_2 \\
\sigma_2 & 0
\end{pmatrix},\quad \tau_2 \sigma_3 = \begin{pmatrix}
0 & -i \sigma_3 \\
i\sigma_3 & 0
\end{pmatrix},\quad \tau_3\sigma_1 = \begin{pmatrix}
\sigma_1 & 0 \\
0 & -\sigma_1
\end{pmatrix}, \quad \tau_0\sigma_2 = \begin{pmatrix}
\sigma_2 & 0 \\
0 & \sigma_2
\end{pmatrix}, \quad \tau_3\sigma_0 = \begin{pmatrix}
\sigma_0 & 0 \\
0 & -\sigma_0
\end{pmatrix}.
\end{equation}
In this way, for example, the matrices $\hat{A}_x$, $\hat{A}_y$, and $\hat{E}$ are shown explicitly as follows:
\begin{align}
\check{A}_x &= \tau_3 A_x - \sigma_2\alpha \equiv \tau_3 \sigma_0 A_x - \tau_0\sigma_2\alpha = \begin{pmatrix}
A_x - \sigma_2\alpha & 0 \\
0 & -A_x - \sigma_2\alpha
\end{pmatrix}, \\
\check{A}_y &= \tau_3 A_y - \sigma_3\alpha  \equiv \tau_3 \sigma_0 A_y - \tau_0\sigma_3\alpha = \begin{pmatrix}
A_y - \sigma_3\alpha & 0 \\
0 & -A_y - \sigma_3\alpha
\end{pmatrix}, \\
&~\check{E} = \tau_3 \eps + \tau_3\sigma_3 h \equiv \tau_3 \sigma_0 \eps - \tau_3\sigma_3h = \begin{pmatrix}
\eps + \sigma_3h & 0 \\
0 & -\eps - \sigma_3 h
\end{pmatrix}. 
\end{align}

\end{widetext}

\subsection{Deriviation of Eqs. (\ref{Sgp2}), (\ref{Snr3}), and (\ref{Cooper_int})}

To obtain the total action in Sec. \ref{Model}, we integrate out the gap fluctuation that deviates from the helical state $\bm{\Delta}^{\rm Helical}_{\rm cl(q)}(\bm{r},t)\equiv\Delta_{\rm cl(q)}\exp\,(i\bm{P}\cdot\bm{r})$. Before doing this, it is convenient to perform the U(1) transformation $Q \rightarrow \check{U}_{P}Q\check{U}^{-1}_P$ on Eqs.~(\ref{Snr}) and (\ref{Sp}), where 
\begin{equation}
\check{U}_P= \begin{pmatrix}
\sigma_0e^{i\bm{P}\cdot\bm{r}/2} & 0 \\
0 &\sigma_0e^{-i\bm{P}\cdot\bm{r}/2}
\end{pmatrix}.
\end{equation}
This U(1) transformation accompanies the replacement $\bm{\Omega}\rightarrow \bm{\Omega}_P\equiv\check{U}^{-1}_P\bm{\Omega}\check{U}_P$ and $\check{\nabla}_i\rightarrow \check{U}^{-1}_P\check{\nabla}_i\check{U}_P$, where the latter results in the substitution of $A^{\rm total}_{i} = A_i+P_i/2$ for the former $A_i$, as described in the text over Eq. (\ref{S1}). Rewriting $A^{\rm total}_i$ as $A_i$ and performing the U(1) transformation, $\tilde{S}_{\rm T}$ remains unchanged, while $\tilde{S}_{\rm N}$ becomes
\begin{equation}
\tilde{S}_{\rm N} = \frac{i\pi \nu}{8} \mathrm{Tr}\lrb{ D(\gamma_{\rm cl}\check{\nabla}_i Q)^2 + 4i \bm{\Omega}_P Q }. \label{Snr2}
\end{equation}
Next, let us extract the fluctuation part from $\bm{\Omega}_P$. Let $\bm{\Omega}^{\rm Helical} \equiv  \gamma_{\rm cl} \check{E} + \gamma_{\rm cl}\check{\bm{\Delta}}^{\rm Helical}_{\rm cl} + \gamma_{\rm q}\check{\bm{\Delta}}^{\rm Helical}_{\rm q}$. Then, 
\begin{align}
\bm{\Omega}_P &= \check{U}^{-1}_P\big(\bm{\Omega}-\bm{\Omega}^{\rm Helical}\big)\check{U}_P + \check{U}^{-1}_P\bm{\Omega}^{\rm Helical}\check{U}_P \notag \\
&= \gamma_{\rm cl}\delta\check{\bm{\Delta}}_{\rm cl}+\gamma_{\rm q}\delta\check{\bm{\Delta}}_{\rm q} + \Omega.
\end{align} 
Here, $\delta\check{\bm{\Delta}}_{\rm cl (q)} = \check{U}^{-1}_P\big(\check{\bm{\Delta}}_{\rm cl (q)} -  \check{\bm{\Delta}}^{\rm Helical}_{\rm cl (q)}\big) \check{U}_P$, $\Omega \equiv \gamma_{\rm cl} \check{E} + \gamma_{\rm cl}\check{\Delta}_{\rm cl} + \gamma_{\rm q}\check{\Delta}_{\rm q}$, and $\check{\Delta}_{\rm cl(q)}=i\tau_2\Delta_{\rm cl (q)}$. From Eq.~(\ref{Snr2}), we then obtain $\tilde{S}_{\rm N}[\bm{\Delta},Q] = {S}_{\rm N}[\Delta,Q] +  \delta S_{\rm N}[\delta\bm{\Delta},Q]$, where $S_{\rm N}$ is exactly Eq. (\ref{Snr3}) and $\delta S_{\rm N}$ is expressed as
\begin{align}
\delta S_{\rm N} &= -\frac{\pi \nu}{2} \mathrm{Tr}\lrb{  \lr{ \gamma_{\rm cl}\delta\check{\bm{\Delta}}_{\rm cl}+\gamma_{\rm q}\delta\check{\bm{\Delta}}_{\rm q} } Q }.\label{dSn}
\end{align}
Now, we integrate out $\exp\,\{ i(\tilde{S}_{\rm HS} + \delta S_{\rm N})\}$ over $\delta\bm{\Delta}_{\rm cl (q)}$ under constraint $\int dtd\bm{r} \, \delta\bm{\Delta}_{\rm cl (q)}=0$, as in Ref.~\cite{Andriyakhina2}. The calculation is elementary, and the necessary term of the result is in the form $\exp\,\{ i(S_{\rm HS}[\Delta] + S^{\rm (c)}_{\rm int}[Q])\}$, where $S_{\rm HS}$ and $S^{\rm (c)}_{\rm int}$ are exactly Eqs. (\ref{Sgp2}) and (\ref{Cooper_int}), respectively.

\subsection{Remark on Eq. (\ref{Cooper_int})}

Actually, the Cooper interaction $S^{\rm (c)}_{\rm int}$, namely Eq. (\ref{Cooper_int}), derived from the procedure presented in the previous section, is slightly different from that derived in the replica formalism \cite{Burmistrov2}. In that work, the condition (in our language) $\int dtd\bm{r} \, \delta\bm{\Delta}_{\rm cl (q)}=0$ is not imposed, resulting in an additional fluctuation. In that case, within our formalism, $Q$ in Eq.~(\ref{Cooper_int}) will be replaced by $Q-\bar{Q}$, where $\bar{Q}$ is the spatio-temporal avarage of $Q$. However, the additional term contributed by $\bar{Q}$ is solely a non-extensive contribution that disappears in the thermodynamic limit due to the contraction of diffusive modes \cite{Burmistrov2}. This is also the case in our formalism, and thus, imposing the condition $\int dtd\bm{r} \, \delta\bm{\Delta}_{\rm cl (q)}=0$ is justified.

\section*{Appendix B: Gaussian Expansion}

\label{AppendixB}
\renewcommand{\theequation}{B.\arabic{equation}}

\setcounter{equation}{0}
\def\thesubsection{\arabic{subsection}}

As described in Sec. \ref{MethodA}, Eq.~(\ref{S_total_r}) can be approximated as $S_{\rm tot}[\Delta,\theta,W] \simeq S_0[W] + S_{\rm MF}[\theta,\Delta] + \tilde{S}_{\rm int}[\theta,W] + S_{\rm I}[\theta,W]$. Each term is obtained by substituting $Q=U\tilde{Q}U^{-1}$ and $\tilde{Q} = \Lambda_{\rm N} + \frac{1}{2}\, [W,\Lambda_{\rm N}] + \frac{1}{8} \, [W,[W,\Lambda_{\rm N}]]$ into the total action $S_{\rm total}$. See the definition of $U$ below Eq. (\ref{spectral1}). The results are shown as follows. First, the terms $S_0$ and $S_{\rm MF}$ are expressed as
\begin{align}     
S_{0} &= \frac{i\pi\nu}{32} \mathrm{Tr} \Big\{ \!D \, \check{\nabla}_i[W,\Lambda_{\rm N}]  \check{\nabla}_i[W,\Lambda_{\rm N}] +  2i\check{E} \, [W,[W,\Lambda_{\rm N}]] \notag \\
&\qquad - D\beta \check{F}_{xy}\, \Lambda_{\rm N}  [\check{\nabla}_x[W,\Lambda_{\rm N}] ,\check{\nabla}_y [W,\Lambda_{\rm N}]]    \!   \Big\}, \label{S0}
\end{align}\vspace{-0.5cm}
\begin{align}
S_{\rm MF}  &=     \frac{i\pi\nu}{8} \mathrm{Tr} \, \Big\{ D[A^{\theta}_i,\Lambda_{\rm N}]^2 + 4i \Omega_{\rm s} \Lambda_{\rm N}  \notag \\
&\qquad - D\beta F^{\theta}_{xy} \Lambda_{\rm N}     [ [A^{\theta}_x,\Lambda_{\rm N}],  [A^{\theta}_y,\Lambda_{\rm N}]  ]     \Big\},
\end{align}
where $\beta=\tau/m$, $A^{\theta}_i =  U_{\rm s}^{-1}\check{\nabla}_i U_{\rm s}  = -i\,( U_{\rm s}^{-1}\check{A}_iU_{\rm s} - \check{A}_i )$,~$\check{F}^{\theta}_{xy} = U_{\rm s}^{-1}\check{F}_{xy} U_{\rm s}$, and $\Omega_{\rm s} = U^{-1}\Omega\,U$ with $\Omega$ given by Eq. (\ref{Esu}). Next, the term $S_{\rm I}$ contains $W$ up to the first order and can be separated into two parts, $S_{\rm I}[\theta,Q] = S_{\rm I1}[\theta,W] + S_{\rm I2}[\theta,W]$, where
\begin{align}
  S_{\rm I1} &=   \frac{i\pi\nu D}{8}\mathrm{Tr}\Big\{  [A^{\theta}_i,\Lambda_{\rm N}][A^{\theta}_i,[W,\Lambda_{\rm N}]] \notag \\
  &\qquad+ 2A^{\theta}_i\Lambda_{\rm N}\check{\nabla}[W,\Lambda_{\rm N}]     \Big\},       \\
  S_{\rm I2} &=\frac{i\pi \nu D \beta }{16} \mathrm{Tr} \, \Big\{ ( [A^{\theta}_y -i\check{A}_y, [[A^{\theta}_x,\Lambda_{\rm N}],  F^{\theta}_{xy}\Lambda_{\rm N}  ]] \notag \\
  &\qquad -  [A^{\theta}_x - i\check{A}_x, [[A^{\theta}_y,\Lambda_{\rm N}],  F^{\theta}_{xy}\Lambda_{\rm N}  ]]  \notag \\
  &\qquad -  [[A^{\theta}_x,\Lambda_{\rm N}], [A^{\theta}_y,\Lambda_{\rm N}]]\, F^{\theta}_{xy} ) \, [W,\Lambda_{\rm N}]    \Big\}.
\end{align}
Lastly, the interaction term is given as $\tilde{S}_{\rm int}[\theta,W] = \tilde{S}^{({\rm c})}_{\rm int}[\theta,W] + \tilde{S}^{\rm (s)}_{\rm int}[\theta,W]$, where
\begin{align}
 \tilde{S}^{\rm (c)}_{\rm int} &=   - \frac{\pi^2\nu\Gamma_{\rm c}}{64} \sum_{i=1,2}  \int dt d\bm{r} \, \mathrm{tr} \lrb{\tau^{\theta}_{\mathrm{cl}i} [W(\bm{r}),\Lambda_{\rm N}]  }_{tt} \notag \\
 &\qquad\qquad\qquad \times \mathrm{tr}\lrb{\tau^{\theta}_{\mathrm{q}i} [W(\bm{r}),\Lambda_{\rm N}]}_{tt},   \label{Sw_intc} \\
 \tilde{S}^{\rm (s)}_{\rm int} &=   - \frac{\pi^2\nu\Gamma_{\rm s}}{64}  \int dt d\bm{r} \,  \mathrm{tr} \lrb{\tau^{\theta}_{\mathrm{cl}0} [W(\bm{r}),\Lambda_{\rm N}]  }_{tt}  \notag \\
 &\qquad\qquad\qquad\times\mathrm{tr}\lrb{\tau^{\theta}_{\mathrm{q}0} [W(\bm{r}),\Lambda_{\rm N}]}_{tt}.       \label{Sw_ints}
\end{align}
Here, $\tau^{\theta}_{\mathrm{cl}\mu}=U^{-1} \gamma_{\rm cl}\tau_{\mu}U,~\tau^{\theta}_{\mathrm{q}\mu} =U^{-1}\gamma_{\rm q}\tau_{\mu}U$, and $\mu=0,1,2,3$. The term like $\{\tau^{\theta}_{\mathrm{cl}\mu} [W(\bm{r}),\Lambda_{\rm N}]  \}_{tt}$ can be represented as
\begin{align}
 \lrb{\tau^{\theta}_{\lambda\mu} [W(\bm{r}),\Lambda_{\rm N}]  }_{tt} &= \frac{1}{\mathcal{T}}\sum_{\eps_1\eps_2}U^{-1}_{\eps_2} \gamma_{\lambda}\tau_{\mu}U_{\eps_1} \notag \\
 &~\times[W_{\eps_1\eps_2}(\bm{r}),\Lambda_{\rm N}]\,e^{-it(\eps_1-\eps_2)}, \label{twl}
 \end{align}
where $\lambda=\mathrm{cl}, \mathrm{q}$. Note that $U^{-1}_{\mathrm{K},\eps}=U_{\mathrm{K}, \eps}$ and
\begin{align}
&~~~U_{\mathrm{K},\eps_2}\gamma_{\rm cl}U_{\mathrm{K},\eps_1} = \begin{pmatrix}
\tau_{\mu} & (F_{\eps_1} - F_{\eps_2})\tau_{\mu} \\
0 & \tau_{\mu}
\end{pmatrix}, \label{Ukl1} \\
& U_{\mathrm{K},\eps_2}\gamma_{\rm q}U_{\mathrm{K},\eps_1} = \begin{pmatrix}
F_{\eps_2}\tau_{\mu} & (F_{\eps_2} F_{\eps_1}-1)\tau_{\mu} \\
-\tau_{\mu} & -F_{\eps_1}\tau_{\mu}
\end{pmatrix}. \label{Ukl2}
\end{align}

\begin{widetext}

\section*{Appendix C: Free Propagators}
\label{AppendixC}
\renewcommand{\theequation}{C.\arabic{equation}}

\setcounter{equation}{0}
\setcounter{subsection}{0}
\def\thesubsection{\arabic{subsection}}

With the help of \textit{wxMaxima}, a computer linear algebra system, $\mathcal{D}^{-1}_{\rm cR(A)}$ and $\mathcal{D}^{-1}_{\rm d}$ are calculated as follows:
\begin{equation}   
\mathcal{D}^{-1}_{\rm cR(A)}(\tilde{\bm{q}},\eps+\eps') \!=\! \begin{pmatrix}
D\tilde{\bm{q}}^2 \!\mp\!i(\eps\!+\!\eps') & 0 &  \pm 4i  D\beta\alpha^3 \tilde{q}_x & \pm 4i ( D\beta\alpha^3 \tilde{q}_y\!+\!\frac{h}{2}) \\
\\
0 & D(\tilde{\bm{q}}^2\!+\!8\alpha^2) \!\mp\! i(\eps\!+\!\eps')  & -4 D\alpha \tilde{q}_y & 4 D\alpha \tilde{q}_x  \\
\\
\pm4i D\beta\alpha^3 \tilde{q}_x &   -4 D\alpha \tilde{q}_y  &  D(\tilde{\bm{q}}^2\!+\!4\alpha^2) \!\mp\! i(\eps\!+\!\eps') & 0 \\
\\
\pm 4i (D\beta\alpha^3 \tilde{q}_y\!+\!\frac{h}{2}) &  4 D\alpha \tilde{q}_x  & 0 & D(\tilde{\bm{q}}^2\!+\!4\alpha^2)\!\mp\!i(\eps\!+\!\eps') 
 \end{pmatrix},    \label{Cooperon}     
 \end{equation}
\begin{equation}   
\mathcal{D}^{-1}_{\rm d}(\tilde{\bm{q}},\eps-\eps') \!=\! \begin{pmatrix}
D\bm{q}^2 \!-\!i(\eps\!-\!\eps') & 0 &  4i  D\beta\alpha^3 q_x &  4i  D\beta\alpha^3 q_y \\
\\
0 & D(\bm{q}^2\!+\!8\alpha^2) \!-\! i(\eps\!-\!\eps')  & -4 D\alpha q_y & 4 D\alpha q_x  \\
\\
4i D\beta\alpha^3 q_x &   -4 D\alpha q_y  &  D(\bm{q}^2\!+\!4\alpha^2) \!-\! i(\eps\!-\!\eps') & 0 \\
\\
4i D\beta\alpha^3 q_y &  4 D\alpha q_x  & 0 & D(\bm{q}^2\!+\!4\alpha^2)\!-\!i(\eps\!-\!\eps') 
 \end{pmatrix}.     \label{Diffuson}
\end{equation}
Here, $C^{\rm t}$ is the transpose matrix of $C$ only for the matrix elements, $C^{\dagger} = \big(C^*)^{\rm t}$, and $C^*$ is the complex conjugation of $C$. Note that there is no gauge-field dependence in the diffuson parts, as mentioned in Sec. \ref{MethodB}. The bare propagators are then obtained as follows:
\begin{align}
\braket{c^{\rm R(A)}_{i\eps_1\eps_2}(\bm{q}_1)c^{\rm R(A)}_{j\eps_3\eps_4}(\bm{q}_2)}_0 &= \frac{2}{\pi\nu} \delta_{\bm{q}_1,-\bm{q}_2}  \delta_{\eps_1\eps_4}\delta_{\eps_2\eps_3} [\mathcal{D}_{\rm cR(A)}(\tilde{\bm{q}}_1,\eps_1+\eps_2)]_{ij}, \\
\braket{b^{\rm U(D)}_{i\eps_1\eps_2}(\bm{q}_1)b^{\rm U(D)}_{j\eps_3\eps_4}(\bm{q}_2)}_0 &= \frac{2}{\pi\nu} \delta_{\bm{q}_1,-\bm{q}_2}  \delta_{\eps_1\eps_4}\delta_{\eps_2\eps_3}  [\mathcal{D}_{\rm d}(\bm{q}_1,\eps_1-\eps_2)]_{ij}. 
 \end{align}

To proceed further, it is convenient to define the following quantities: 
\begin{align}
K^{\rm s\pm}_{\mathrm{cR(A)},\eps\eps'} (\bm{q}) &\equiv K^{00\pm}_{\mathrm{cR(A)},\eps\eps'} (\bm{q}) + K^{33\pm}_{\mathrm{cR(A)},\eps\eps'} (\bm{q}), \label{Ks}\\
 K^{\mathrm{a}}_{ \mathrm{cR(A)},\eps\eps'} (\bm{q}) \equiv K^{00+}_{\mathrm{cR(A)},\eps\eps'}  (\bm{q})& - K^{33+}_{\mathrm{cR(A)},\eps\eps'}  (\bm{q})=  K^{00-}_{\mathrm{cR(A)},\eps\eps'} (\bm{q}) - K^{33-}_{\mathrm{cR(A)},\eps\eps'} (\bm{q}), \label{Ka} \\
 K^{\pm}_{\mathrm{d},\eps\eps'} (\bm{q}) &\equiv K^{00}_{\mathrm{d},\eps\eps'} (\bm{q}) \pm K^{33}_{\mathrm{d},\eps\eps'} (\bm{q}),  \label{Kd}
\end{align}
where $K^{00\pm}_{\rm cR(A)} = \frac{2}{\pi\nu} \{ [\mathcal{D}_{\rm cR(A)}]_{00} \pm [\mathcal{D}_{\rm cR(A)}]_{03}\} $, $K^{33\pm}_{\rm cR(A)} =\frac{2}{\pi\nu} \{ [\mathcal{D}_{\rm cR(A)}]_{33} \pm [\mathcal{D}_{\rm cR(A)}]_{03} \}$, and $K^{ii}_{\rm d} =\frac{2}{\pi\nu}  [\mathcal{D}_{\rm d}]_{ii}$ (the dependence on $\eps,~\eps'$, and $\bm{q}$ is abbreviated).

\section*{Appendix D: Derivation of Modified Usadel equation}
\label{AppendixD}
\renewcommand{\theequation}{D.\arabic{equation}}

\setcounter{equation}{0}
\setcounter{subsection}{0}
\def\thesubsection{\arabic{subsection}}

\subsection{Argument of the divergent terms\label{AppendixD1}}

In the following, we present the details of the derivation of the modified Usadel equation. We first explain how we came to the ansatz Eq.~(\ref{spectralci}). The problem of unphysical divergence comes from the interaction term $\tilde{S}_{\rm int}$. We first adjust $\tilde{S}_{\rm int}$ in a way that allows us to perform the concrete calculation. From Eqs.~(\ref{twl}), (\ref{Ukl1}), (\ref{Ukl2}), (\ref{CR}), and (\ref{BB}), we obtain

\begin{align}
  \mathrm{tr} \lrb{\tau^{\theta}_{\mathrm{cl}\mu} [W(\bm{r}),\Lambda_{\rm N}]  }_{tt}  = \frac{2}{\mathcal{T}} \sum_{\eps_{1}\eps_{2}} \mathrm{tr}\,\big\{ & \tau_3\big[ \check{U}^{-1}_{\mathrm{sR},\eps_2} \tau_{\mu}  \check{U}_{\mathrm{sR},\eps_1}\check{C}_{\mathrm{R},\eps_1\eps_2}  - \check{U}^{-1}_{\mathrm{sA},\eps_2} \tau_{\mu} \check{U}_{\mathrm{sA},\eps_1}\check{C}_{\mathrm{A},\eps_1\eps_2}  \notag  \\
  &+     (F_{\eps_2}-F_{\eps_1})\,\check{U}^{-1}_{\mathrm{sR},\eps_2} \tau_{\mu} \check{U}_{\mathrm{sA},\eps_1}\check{\bar{B}}_{\eps_1\eps_2}       \big]   \big\} \,e^{-it(\eps_1-\eps_2)}, \\
  \mathrm{tr} \lrb{\tau^{\theta}_{\mathrm{q}\mu} [W(\bm{r}),\Lambda_{\rm N}]}_{tt} = \frac{2}{\mathcal{T}} \sum_{\eps_{3}\eps_4} \mathrm{tr}\,\big\{ & \tau_3\big[ F_{\eps_4}\check{U}^{-1}_{\mathrm{sR},\eps_4} \tau_{\mu}  \check{U}_{\mathrm{sR},\eps_3}\check{C}_{\mathrm{R},\eps_3\eps_4} + F_{\eps_3}\check{U}^{-1}_{\mathrm{sA},\eps_4} \tau_{\mu} \check{U}_{\mathrm{sA},\eps_3}\check{C}_{\mathrm{A},\eps_3\eps_4}  \notag  \\
  &+     (F_{\eps_4}F_{\eps_3}-1)\,\check{U}^{-1}_{\mathrm{sR},\eps_4} \tau_{\mu} \check{U}_{\mathrm{sA},\eps_3}\check{\bar{B}}_{\eps_3\eps_4}  +  \check{U}^{-1}_{\mathrm{sA},\eps_4} \tau_{\mu} \check{U}_{\mathrm{sR},\eps_3}\check{B}_{\eps_3\eps_4}     \big]   \big\} \,e^{-it(\eps_3-\eps_4)}. 
\end{align}
Multiplying these terms together as one equation and averaging it with $S_0$, we obtain
\begin{equation}   
\braket{\tilde{S}_{\rm int}}_0 = -\frac{\pi^2\nu}{16} \lra{\Gamma_{\rm s} \mathcal{S}_{\rm I 0}   +  \Gamma_{\rm c} \lr{ \mathcal{S}_{\rm I1} + \mathcal{S}_{ \rm I2} }},   
\end{equation}
\begin{align}
\mathcal{S}_{\mathrm{I}i} = \frac{1}{\mathcal{T}} \, \textstyle{ \sum_{\eps_1,\eps_2}}  \int d\bm{r}\,\displaystyle \Big[ F_{\eps_1} &\mathrm{tr}\{  \tau_3 \check{U}^{-1}_{\mathrm{sR},\eps_2} \tau_i \check{U}_{\mathrm{sR},\eps_1}  \sharp\check{C}_{\mathrm{R},\eps_1\eps_2}   \} \, \mathrm{tr}\{ \tau_3 \check{U}^{-1}_{\mathrm{sR},\eps_1} \tau_i \check{U}_{\mathrm{sR},\eps_2}  \sharp \check{C}_{\mathrm{R},\eps_2\eps_1} \}    \notag  \\
  &-F_{\eps_2} \mathrm{tr}\{  \tau_3 \check{U}^{-1}_{\mathrm{sA},\eps_2} \tau_i \check{U}_{\mathrm{sA},\eps_1}  \sharp\check{C}_{\mathrm{A},\eps_1\eps_2}   \} \, \mathrm{tr}\{ \tau_3 \check{U}^{-1}_{\mathrm{sA},\eps_1} \tau_i \check{U}_{\mathrm{sA},\eps_2}  \sharp \check{C}_{\mathrm{A},\eps_2\eps_1} \}   \notag   \\
  &-\lr{F_{\eps_1}-F_{\eps_2}} \mathrm{tr}\{  \tau_3 \check{U}^{-1}_{\mathrm{sA},\eps_2} \tau_i \check{U}_{\mathrm{sR},\eps_1}  \sharp\check{B}_{\eps_1\eps_2}   \} \, \mathrm{tr}\{ \tau_3 \check{U}^{-1}_{\mathrm{sR},\eps_1} \tau_i \check{U}_{\mathrm{sA},\eps_2}  \sharp \check{\bar{B}}_{\eps_2\eps_1} \}    \Big],     \label{concrete_int}
   \end{align}
where the sharp symbol $\sharp$ between fields indicates the Wick contraction with respect to $S_0$. Note that the contractions of the pair $\check{C}_{\rm R}$-$\check{B}$ or $\check{\bar{B}}$, $\check{C}_{\rm A}$-$\check{B}$ or $\check{\bar{B}}$, and the pair $\check{C}_{\rm R}$-$\check{C}_{\rm A}$ become zero. 

The concrete calculation of the above equation will be performed in the section. Here, let us point out the divergence problem and how to solve it. In the above equation, for example, there are the terms 
\begin{align}
\Delta_{\rm cl} \sum_{\bm{q},\eps,\eps'} & \lr{F_{\eps}+F_{\eps'}} \Big\{   K^{\rm s\pm}_{\mathrm{cR},\eps\eps'} (\bm{q}) \big(  f^{\rm R}_{1\eps' \pm} - f^{\rm R}_{1\eps \pm} \big)   F_{\eps}  \theta^{\rm Rq}_{1\eps\pm} +   K^{\rm s\mp}_{\mathrm{cA},\eps\eps'} (\bm{q}) \big(  f^{\rm A}_{1\eps' \mp} - f^{\rm A}_{1\eps \mp} \big)  F_{\eps}  \theta^{\rm Aq}_{1\eps\mp} \Big\}  . \label{Idv}
\end{align}
First, the terms in the form $ F^2_{\eps} K^{\rm s\pm}_{\mathrm{cR},\eps\eps'}f^{\rm R}_{1\eps'\pm}\theta^{\rm Rq}_{1\eps\pm}$ vanish due to the integration over $\eps'$ because it is the integration of the multiplication between two retarded functions \cite{Kamenevhon}. We then neglect those terms and consider the functional differentiation with respect to $\theta^{\rm Rq}_{\eps\pm}$ of the remaining terms. If the retarded and advanced spectral functions were independent,  the result would become in the following:
\begin{align}
\Delta_{\rm cl} F_{\eps} \sum_{\bm{q},\eps'}  F_{\eps'}  K^{\rm s\pm}_{\mathrm{cR},\eps\eps'} (\bm{q})  f^{\rm R}_{1\eps' \pm}  - \Delta_{\rm cl} F_{\eps} \sum_{\bm{q},\eps'}  \lr{F_{\eps}+F_{\eps'}}   K^{\rm s\pm}_{\mathrm{cR},\eps\eps'} (\bm{q})  f^{\rm R}_{1\eps \pm} .   \label{testue}
\end{align}
However, these terms are actually divergent due to the second term. To eliminate it, the advanced counterpart is necessary. In this way, the retarded and advanced spectral functions cannot be considered independent of each other. 

We then reconsider Eq.~(\ref{Idv}) by focusing on the summation
\begin{align}
\sum_{\bm{q},\eps,\eps'} F_{\eps} \lr{F_{\eps}+F_{\eps'}}  \Big\{ K^{\rm s\pm}_{\mathrm{cR},\eps\eps'} (\bm{q})  f^{\rm R}_{1\eps \pm}  \theta^{\rm Rq}_{1\eps\pm} + K^{\rm s\mp}_{\mathrm{cA},\eps\eps'} (\bm{q})  f^{\rm A}_{1\eps \mp}  \theta^{\rm Aq}_{1\eps\mp} \Big\}.  \label{ldd}
\end{align}
To deduce the relation between retarded and advanced spectral functions, we look at the special case that neglects the interactions and the Rashba SOC. In this case, when performing the functional derivative of the action with respect to $\theta^{\rm Rq}_{1\eps\pm}$, the term with momentum and energy integration like Eq.~(\ref{Idv}) does not appear, and the spectral functions can be solved analytically. It can further be shown that their solutions satisfy the ansatz of Eq.~(\ref{spectralci}). Moreover, for propagators of the cooperon modes, it appears that
\begin{align}
K^{\rm s\pm}_{\mathrm{cR},\eps\eps'}(\bm{q}) = K^{\rm s\mp}_{\mathrm{cA},-\eps,-\eps'}(-\bm{q}),  \label{Kss} \\
K^{\rm a}_{\mathrm{cR},\eps\eps'}(\bm{q}) = K^{\rm a}_{\mathrm{cA},-\eps,-\eps'}(-\bm{q}), \label{Kaa}
\end{align}
which can be shown by replacing $c^{\rm A}_{1,\eps\eps'}(-\bm{q})$ with $-c^{\rm A}_{1,\eps\eps'}(-\bm{q})$ in Eq.~(\ref{Seff_Cooperon}), and the propagators of the relevant components remain invariant. Using the ansatz of Eq.~(\ref{spectralci}) and Eqs. (\ref{Kss}) and (\ref{Kaa}), we see that Eq.~(\ref{ldd}) vanishes.

The terms like Eq.~(\ref{ldd}), in which the classical parts of the spectral functions have the same energy dependence as the quantum parts, also appear in the equations at the $\Delta^3_{\rm cl}$-order. They cancel out by the same argument as above. Note that the terms with multiplication between more than two retarded or advanced functions, such as $F^2_{\eps}K^{\rm s\pm}_{\mathrm{cR},\eps\eps'}f^{\rm R}_{1\eps'\pm}\theta^{\rm Rq}_{1\eps\pm}$ considered above, also vanish by energy integration.

\subsection{Calculations for deriving the modified Usadel equation}

We can now calculate Eq.~(\ref{UsadelS}). The calculation is performed based on the ansatz of Eq.~(\ref{spectralci}) and Eqs.~(\ref{Kss}) and (\ref{Kaa}). First, the terms associated with the ordinary mean-field terms $S_{\rm MF}$ are calculated with the help from \textit{wxMaxima} as follows:
\begin{align}
\frac{1}{2F_{\eps}\mathcal{A}}&\Bigg[\frac{\delta}{\delta\theta^{\rm qR}_{\eps\pm }} \lra{ \mathrm{Tr}\, \{ \Omega_{\rm s}\Lambda_{\rm N}\} } \Bigg]_{\theta^{\rm q}=0} = 2\Delta_{\rm cl} ( \eps_{\pm} f^{\rm R}_{1\eps\pm} - 1) + \frac{\Delta^3_{\rm cl}}{3} \lr{ 6\eps_{\pm} f^{\rm R}_{3\eps\pm} -3[f^{\rm R}_{1\eps\pm}]^2 + \eps_{\pm}[f^{\rm R}_{1\eps\pm}]^3 }, \label{MF_N} \\
\frac{1}{2F_{\eps}\mathcal{A} }\Bigg[ \frac{\delta}{\delta \theta^{\rm qR}_{\eps \pm}}&\lra{ \mathrm{Tr}\lrb{  [A^{\theta}_{i},\Lambda_{\rm N}]^2  - \beta F^{\theta}_{xy} \Lambda_{\rm N}     [ [A^{\theta}_x,\Lambda_{\rm N}],  [A^{\theta}_y,\Lambda_{\rm N}]  ]    } } \Bigg]_{\theta^{\rm q}=0} \notag \\
 &= 8\Delta_{\rm cl} \lr{ \alpha^2 f^{\rm R}_{1\eps \mp} - [\alpha^2 + 2A_y^2 \pm 4i\alpha^3\beta A_y]  f^{\rm R}_{1\eps \pm}  } \notag \\
 &\qquad + \frac{1}{3}\Delta^3_{\rm cl} \Big(       24\alpha^2 f^{\rm R}_{3\eps \mp} + [4\alpha^2 \mp 24 i \beta\alpha^3 A_y] [f^{\rm R}_{1\eps \mp}]^3 -12\alpha^2 f^{\rm R}_{1\eps \pm}[f^{\rm R}_{1\eps \mp}]^2 + 12\alpha^2 \,[ 1 \pm 6i\beta\alpha] [f^{\rm R}_{1\eps \pm}]^2f^{\rm R}_{1\eps \mp} \notag \\
 &\qquad - 24\, [\alpha^2 + 4A^2_y \pm 4i\beta\alpha^3 A_y]f^{\rm R}_{3\eps \pm} - 4 \, [\alpha^2 + 8A^2_y \pm 16i\beta\alpha^3A_y][f^{\rm R}_{1\eps \pm}]^3 \Big), \label{MF_R}
\end{align}
where the factor 2 in front of $F_{\eps}$ comes from the advanced part which is not actually independent of the retarded part, as discussed in Appendix~\hyperref[AppendixD1]{D1}. Thus, differentiating with respect to $\theta^{\rm Rq}_{\eps\pm}$ also affects the advanced part.

For the kinetic term $S^2_{\rm I}$, we need to be careful to the Wick contraction. The calculation using \textit{wxMaxima} may be easily performed after considering the following. First, note that $\braket{S^2_{\rm I}}_0$ is in the form $\braket{\mathrm{Tr}\,\{PW\}\mathrm{Tr}\,\{PW\}}_0$ for a certain $P$. When performing each trace only in matrix elements, the term $\mathrm{Tr}\,\{PW\}$ can be transformed into the form $G^{\rm R(A)}\Phi^{\rm cR(A)}$, where $G^{\rm R(A)}$ is a certain row vector and $\Phi^{\rm cR(A)}$ is the column vector field of the cooperon modes, defined in Sec.~\ref{MethodB}. In the case of $S^2_{\rm I}$, $G^{\rm R(A)}$ is stationary in time, and the momentum $\bm{q}$ in $\Phi^{\rm cR(A)}$ is zero due to the Fourier transform, i.e., $\tilde{\bm{q}} = A_y\bm{e}_y$. Note that only cooperon modes contribute to this kinetic term. On the other hand, we transform the other term to be $[\Phi^{\rm cR(A)}]^{\rm t}[G^{\rm R(A)}]^{\rm t}$. Gathering them together, we obtain
\begin{align} 
\braket{S^2_{\rm I}}_0 = \sum_{\eps_i} \sum_{\lambda = \mathrm{R,A}} G^{\lambda}_{\eps_1}\braket{\Phi^{\rm c\lambda}_{\eps_1\eps_1}(\bm{q}=0)[\Phi^{\rm c\lambda}_{\eps_2\eps_2}(\bm{q}'=0)]^{\rm t} }_0[G^{\lambda }_{\eps_2}]^{\rm t} = \frac{2}{\pi\nu}\sum_{\eps} \sum_{\lambda = \mathrm{R,A}} G^{\lambda}_{\eps} \big[\mathcal{D}_{\rm c\lambda}(A_y\bm{e}_y,2\eps)\big]\, [G^{\lambda }_{\eps}]^{\rm t}.  \label{testk}
\end{align}
Thus, we have to find only the row vector $G^{\lambda}$ from \textit{wxMaxima}. Ultimately, the result can be written as $S^2_{\rm I} = \Delta_{\rm cl} S^2_{\mathrm{I},\Delta_{\rm cl}} + \Delta^3_{\rm cl} S^2_{\mathrm{I},\Delta^3_{\rm cl}}$, where
\begin{align}
\frac{1}{2F_{\eps} \mathcal{A}}  \lra{\frac{\delta}{\delta \theta^{\rm qR}_{\eps \pm}} \braket{S^2_{\mathrm{I},\Delta_{\rm cl}}}_0 }_{\theta^{\rm q}=0}\! = - 2 \lr{ \frac{\pi\nu D}{4} }^2 \Big[  &  -32A^2_y \tilde{K}^{\rm s\pm}_{\mathrm{cR},\eps\eps}(   \beta^2\alpha^6 + A^2_y   ) f^{\rm R}_{1\eps \pm} +  32A^2_y \tilde{K}^{\rm a}_{\mathrm{cR},\eps\eps} (\beta^2 \alpha^6  -  A^2_y  ) f^{\rm R}_{1\eps \mp}                    \Big],  \label{KI_Delta1} \\
\frac{1}{2F_{\eps} \mathcal{A}} \lra{\frac{\delta}{\delta \theta^{\rm qR}_{\eps \pm}} \braket{S^2_{\mathrm{I},\Delta^3_{\rm cl} }}_0 }_{\theta^{\rm q}=0} \!= - 2 \lr{ \frac{\pi\nu D}{4} }^2 \Big[ &  \Lambda^{\rm R\pm}_{1\eps} f^{\rm R}_{3\eps \mp}  + \Lambda^{\rm R\pm}_{2\eps} [f^{\rm R}_{1\eps \mp}]^3 + \Lambda^{\rm R\pm}_{3\eps} f^{\rm R}_{1\eps\pm} [f^{\rm R}_{1\eps \mp}]^2 + \Lambda^{\rm R\pm}_{4\eps} [f^{\rm R}_{1\eps\pm}]^2 f^{\rm R}_{1\eps \mp} \notag \\
&   + \Lambda^{\rm R\pm}_{5\eps} [f^{\rm R}_{1\eps \pm}]^3 + \Lambda^{\rm R\pm}_{6\eps} f^{\rm R}_{3\eps \pm} \Big], \label{KI_Delta3}
\end{align}
where
\begin{align}
\Lambda^{\rm R\pm}_{1\eps} &= 32A^2_y \tilde{K}^{\rm a}_{\mathrm{cR},\eps\eps} (\beta^2\alpha^6-A^2_y),  \\
\Lambda^{\rm R\pm}_{2\eps} &= \frac{2}{3}A^2_y\beta^2\alpha^6 (14\tilde{K}^{\rm a}_{\mathrm{cR},\eps\eps} + 3\tilde{K}^{\rm s\pm}_{\mathrm{cR},\eps\eps} + 3\tilde{K}^{\rm s\mp}_{\mathrm{cR},\eps\eps}) \mp 2iA_y\beta\alpha^5 (\tilde{K}^{\rm s\pm}_{\mathrm{cR},\eps\eps} - \tilde{K}^{\rm s\mp}_{\mathrm{cR},\eps\eps}) \notag \\ 
&\quad \mp \frac{8i}{3}A^3_y\beta\alpha^3 (-4\tilde{K}^{\rm a}_{\mathrm{cR},\eps\eps} + 3\tilde{K}^{\rm s\pm}_{\mathrm{cR},\eps\eps} + \tilde{K}^{\rm s\mp}_{\mathrm{cR},\eps\eps}) -\frac{64}{3}A^4_y \tilde{K}^{\rm a}_{\mathrm{cR},\eps\eps},    \\
\Lambda^{\rm R\pm}_{3\eps} &= -4A^2_y\beta^2\alpha^6 (3\tilde{K}^{\rm a}_{\mathrm{cR},\eps\eps} + 2\tilde{K}^{\rm s\pm}_{\mathrm{cR},\eps\eps} + \tilde{K}^{\rm s\mp}_{\mathrm{cR},\eps\eps} ) \mp \frac{4i}{3} A_y\beta\alpha^3 (3\alpha^2 + 4A^2_y) (\tilde{K}^{\rm a}_{\mathrm{cR},\eps\eps} + \tilde{K}^{\rm s\mp}_{\mathrm{cR},\eps\eps} - 2 \tilde{K}^{\rm s\pm}_{\mathrm{cR},\eps\eps}), \\
\Lambda^{\rm R\pm}_{4\eps} &= \frac{2}{3}A^2_y\beta^2\alpha^6 (26 \tilde{K}^{\rm a}_{\mathrm{cR},\eps\eps} + 15\tilde{K}^{\rm s\pm}_{\mathrm{cR},\eps\eps} + 3\tilde{K}^{\rm s\mp}_{\mathrm{cR},\eps\eps}) \pm 2iA_y\beta\alpha^5 (4\tilde{K}^{\rm a}_{\mathrm{cR},\eps\eps}  + \tilde{K}^{\rm s\mp}_{\mathrm{cR},\eps\eps} - 5\tilde{K}^{\rm s\pm}_{\mathrm{cR},\eps\eps}) \notag \\
&\qquad \mp \frac{8i}{3} A^3_y\beta\alpha^3 (    8 \tilde{K}^{\rm a}_{\mathrm{cR},\eps\eps} - 5 \tilde{K}^{\rm s\pm}_{\mathrm{cR},\eps\eps} - 3 \tilde{K}^{\rm s\mp}_{\mathrm{cR},\eps\eps}) - \frac{64}{3}A^4_y \tilde{K}^{\rm a}_{\mathrm{cR},\eps\eps},   \\ 
\Lambda^{\rm R\pm}_{5\eps} &= \frac{4}{3} A^2_y\beta^2\alpha^6 (3\tilde{K}^{\rm a}_{\mathrm{cR},\eps\eps} + 11\tilde{K}^{\rm s\pm}_{\mathrm{cR},\eps\eps}) \mp 4iA_y\beta\alpha^3 (\alpha^2-4A^2_y) (\tilde{K}^{\rm a}_{\mathrm{cR},\eps\eps} - \tilde{K}^{\rm s\pm}_{\mathrm{cR},\eps\eps}) - \frac{128}{3}A^4_y \tilde{K}^{\rm s\pm}_{\mathrm{cR},\eps\eps}, \\
\Lambda^{\rm R\pm}_{6\eps} &= -32 A^2_y \tilde{K}^{\rm s\pm}_{\mathrm{cR},\eps\eps} (\beta^2\alpha^6 + A^2_y).
\end{align}
Here $\tilde{K}^{\rm s\pm}_{\mathrm{cR},\eps\eps} = K^{\rm s\pm}_{\mathrm{cR},\eps\eps}(A_y)$ and $\tilde{K}^{\rm a}_{\mathrm{cR},\eps\eps} = K^{\rm a}_{\mathrm{cR},\eps\eps}(A_y)$. In the numerical calculation, we find that the contribution from Eqs.~(\ref{KI_Delta1}) and (\ref{KI_Delta3}) is much smaller than the summation of Eqs.~(\ref{MF_N}) and (\ref{MF_R}). Nevertheless, we include it in the results of this work. Note that only cooperon modes contribute to $S_{\rm MF}$ and $S^2_{\rm I}$.

The method used to calculate the interaction term is the same as that used for the kinetic term. However, unlike the kinetic term, contributions from the diffuson modes and one-loop momentum integration appear when calculating the interaction term. Moreover, the row vector $G^{\lambda}$ in Eq.~(\ref{testk}) becomes dependent on two energies, say $\eps_1$ and $\eps_2$. For example, we have to calculate the terms in the form
\begin{align} 
 \sum_{\eps_i}  \sum_{\bm{q},\bm{q}'} \sum_{\lambda = \mathrm{R,A}} G^{\lambda}_{\eps_2\eps_1}\braket{\Phi^{\rm c\lambda}_{\eps_1\eps_2}(\bm{q})[\Phi^{\rm c\lambda}_{\eps_4\eps_3}(\bm{q}')]^{\rm t} }_0[G^{\lambda }_{\eps_3\eps_4}]^{\rm t} = \frac{2}{\pi\nu} \sum_{\eps,\eps'}\sum_{\bm{q}} \sum_{\lambda = \mathrm{R,A}} G^{\lambda}_{\eps\eps'} \big[\mathcal{D}_{\rm c\lambda}(\bm{q},\eps\!+\!\eps')\big]\, [G^{\lambda }_{\eps'\eps}]^{\rm t}.
\end{align}
The explicit form of $G^{\lambda}$ can be obtained using \textit{wxMaxima} on Eq.~(\ref{concrete_int}). We write the result in the form $\tilde{S}_{\rm int} = \Delta_{\rm cl} S_{\mathrm{int},\Delta_{\rm cl}} + \Delta^3_{\rm cl} S_{\mathrm{int},\Delta^3_{\rm cl}}$. However, the results obtained from performing algebraic operations on matrices using that program cannot be used immediately because we must take into account the argument in Appendix~\hyperref[AppendixD1]{D1}. For example, we demonstrate the result for $S_{\mathrm{int},\Delta_{\rm cl}}$ from the program here. Rather than obtaining $\braket{\tilde{S}_{\rm int}}$ as the final result, it is more convenient to calculate $[\delta\braket{\tilde{S}_{\rm int}}/\delta\theta^{\rm R(A)q}]_{\theta^{\rm q}}=0$ directly as the final output and then neglect the unphysical terms. By ignoring the argument in Appendix~\hyperref[AppendixD1]{D1}, we will reach 
\begin{align}
\frac{16 \mathcal{T} }{F_{\eps}\pi^2\nu }\lra{\frac{\delta}{\delta \theta^{\rm qR}_{\eps\pm}}\braket{ S_{\mathrm{int},\Delta_{\rm cl}} }_0 }_{\theta^{\rm q}=0} \! \overset{!}{=}& - \Gamma_{\rm s} \sum_{\bm{q},\eps'} (F_{\eps'}+F_{\eps}) \Big[ K^{\rm s\pm}_{\mathrm{cR},\eps\eps'}  (\bm{q})  (f^{\rm R}_{1\eps'\pm} - f^{\rm R}_{1\eps\pm} ) + K^{\rm a}_{\mathrm{cR},\eps\eps'}  (\bm{q})  (f^{\rm R}_{1\eps'\mp} - f^{\rm R}_{1\eps\pm} )   \Big] \notag \\
&- \Gamma_{\rm s} \sum_{\bm{q},\eps'} (F_{\eps'}-F_{\eps}) \Big[ \big( K^+_{\mathrm{d},\eps\eps'}  (\bm{q})   + K^-_{\mathrm{d},\eps\eps'} (\bm{q}) \big) (f^{\rm A}_{1\eps'\pm} - f^{\rm A}_{1\eps\pm})  \Big] \notag \\
&-  2\Gamma_{\rm c} \sum_{\bm{q},\eps'} (F_{\eps'}+F_{\eps}) (K^{ \rm s\pm}_{\mathrm{cR},\eps\eps'}  (\bm{q}) + K^{\rm a}_{\mathrm{cR},\eps\eps'}  (\bm{q})) f^{\rm R}_{1\eps\pm} \notag \\
&-  2\Gamma_{\rm c} \sum_{\bm{q},\eps'} (F_{\eps'}-F_{\eps}) \big( K^+_{\mathrm{d},\eps\eps'} f^{\rm R}_{1\eps\pm}  (\bm{q}) + K^-_{\mathrm{d},\eps\eps'} (\bm{q}) f^{\rm R}_{1\eps\mp} \big).       
\end{align}
Many terms in the form of Eq.~(\ref{testue}) appear in this equation. However, such terms must be neglected when considering the argument in Appendix.~\hyperref[AppendixD1]{D1}. Therefore, the correct result is
\begin{equation}
\frac{16 \mathcal{T} }{2F_{\eps}\pi^2\nu }\lra{\frac{\delta}{\delta \theta^{\rm qR}_{\eps\pm}}\braket{ S_{\mathrm{int},\Delta_{\rm cl}} }_0 }_{\theta^{\rm q}=0} \! =  - \Gamma_{\rm s} \sum_{\bm{q},\eps'} F_{\eps'} \Big[ K^{\rm s\pm}_{\mathrm{cR},\eps\eps'}  f^{\rm R}_{1\eps'\pm} + K^{\rm a}_{\mathrm{cR},\eps\eps'} f^{\rm R}_{1\eps'\mp} + \big( K^+_{\mathrm{d},\eps\eps'} + K^-_{\mathrm{d},\eps\eps'} \big) f^{\rm A}_{1\eps'\pm}    \Big].   \label{int_Delta1}     
\end{equation}
Note that the factor 2 in front of $F_{\eps}$ appears for the same reason as described below Eq.~(\ref{MF_R}), and the Cooper channel vanishes. Moreover, we have abbreviated the momentum dependence of propagators. The calculation for the term $S_{\mathrm{int},\Delta^3_{\rm cl}}$ can be performed in the same way. Let $S_{\mathrm{int},\Delta^3_{\rm cl}} = S^{\rm (s)}_{\mathrm{int},\Delta^3_{\rm cl}} + S^{\rm (c)}_{\mathrm{int},\Delta^3_{\rm cl}}$, we obtain the following:

\begin{align}
&\frac{16 \mathcal{T} }{2F_{\eps}\pi^2\nu }\lra{\frac{\delta}{\delta \theta^{\rm qR}_{\eps\pm}}\braket{ S^{\rm (c)}_{\mathrm{int},\Delta^3_{\rm cl}} }_0 }_{\theta^{\rm q}=0} \! =  - \frac{\Gamma_{\rm c}}{4} \sum_{\bm{q},\eps'} F_{\eps'}  \Big\{ 2 K^{\rm a}_{\mathrm{cR}, \eps\eps'} \big( f^{\rm R}_{1\eps\pm}  [f^{\rm R}_{1\eps' \mp}]^2 +  f^{\rm R}_{1\eps\mp} f^{\rm R}_{1\eps'\mp} f^{\rm R}_{1\eps'\pm}  \big) + 4 K^{\rm s\pm}_{\mathrm{cR},\eps\eps'} f^{\rm R}_{1\eps\pm}[f^{\rm R}_{1\eps'\pm}]^2 \big)  \notag \\
&\qquad + K^-_{\mathrm{d},\eps\eps'} \big(   f^{\rm R}_{1\eps \mp}[f^{\rm A}_{1\eps' \mp}]^2 + 2 f^{\rm R}_{1\eps \pm}f^{\rm A}_{1\eps' \pm}f^{\rm A}_{1\eps' \mp} + f^{\rm R}_{1\eps \mp} [f^{\rm A}_{1\eps' \pm}]^2  +  f^{\rm R}_{1\eps \mp} [f^{\rm R}_{1\eps' \pm}]^2 \big) +  4 K^+_{\mathrm{d},\eps\eps'}  f^{\rm R}_{1\eps \pm} [f^{\rm A}_{1\eps' \pm}]^2    \Big\},  \label{intc_Delta3} \\
&\frac{16 \mathcal{T} }{2F_{\eps} \pi^2\nu}\lra{\frac{\delta}{\delta \theta^{\rm qR}_{\eps\pm}}\braket{ S^{\rm (s)}_{\mathrm{int},\Delta^3_{\rm cl}} }_0 }_{\theta^{\rm q}=0} \! = - \frac{\Gamma_{\rm s}}{24} \sum_{\bm{q},\eps'} F_{\eps'}  \Big\{ K^{\rm a}_{\mathrm{cR},\eps\eps'} \big ( 24 f^{\rm R}_{3\eps' \mp} + [f^{\rm R}_{1\eps' \mp}]^3 + 3 [f^{\rm R}_{1\eps \pm}]^2 f^{\rm R}_{1\eps'\mp} - 6 f^{\rm R}_{1\eps \pm}f^{\rm R}_{1\eps' \pm}f^{\rm R}_{1\eps'\mp} \notag \\
&\qquad + 3 [f^{\rm R}_{1\eps\mp}]^2f^{\rm R}_{1\eps' \mp} + 3 f^{\rm R}_{1\eps' \mp}  [f^{\rm R}_{1\eps' \pm}]^2  - 3 f^{\rm R}_{1\eps \mp} [f^{\rm R}_{1\eps' \pm}]^2 + 6 f^{\rm R}_{1\eps \mp} f^{\rm R}_{1\eps \pm}f^{\rm R}_{1\eps' \pm} - 3 f^{\rm R}_{1\eps \mp}[f^{\rm R}_{1\eps' \mp}]^2 \big) \notag\\
&\qquad + 4 K^{\rm s \pm}_{\mathrm{cR},\eps\eps'}  \big( 6 f^{\rm R}_{3\eps' \pm } +  [f^{\rm R}_{1\eps' \pm}]^3 -  3 f^{\rm R}_{1\eps \pm}[f^{\rm R}_{1\eps' \pm}]^2 + 3 [f^{\rm R}_{1\eps\pm}]^2 f^{\rm R}_{1\eps' \pm } \big) \notag \\
&\qquad + K^-_{\mathrm{d},\eps\eps'} \big( 3 f^{\rm A}_{1\eps'\pm} [f^{\rm A}_{1\eps'\mp}]^2 - 6  f^{\rm R}_{1\eps\pm}[f^{\rm A}_{1\eps'\mp}]^2  - 6 f^{\rm R}_{1\eps\mp} f^{\rm A}_{1\eps' \pm} f^{\rm A}_{1\eps' \mp} + 6 f^{\rm R}_{1\eps \pm} f^{\rm R}_{1\eps \mp} f^{\rm A}_{1\eps' \mp} + 3 [f^{\rm R}_{1\eps \mp}]^2f^{\rm A}_{1\eps' \pm} \big) \notag \\
&\qquad  +  \big( 4 K^+_{\mathrm{d},\eps\eps'} + K^-_{\mathrm{d},\eps\eps'} \big) \big([f^{\rm A}_{1\eps' \pm}]^3  + 3\, [f^{\rm R}_{1\eps \pm}]^2f^{\rm A}_{1\eps' \pm}   \big)+ 24 \big( K^+_{\mathrm{d},\eps\eps'} + K^-_{\mathrm{d},\eps\eps'} \big)  f^{\rm A}_{3\eps' \pm} - 12 K^+_{\mathrm{d},\eps\eps'} f^{\rm R}_{1\eps \pm} [f^{\mathrm{A}}_{1\eps' \pm}]^2    \Big\}.  \label{ints_Delta3}
\end{align}
As in Eq.~(\ref{int_Delta1}), the momentum dependence of propagators in these equations is abbreviated.

\subsection{Modified Usadel equation} \label{secIIIC}

We now combine Eqs.~(\ref{MF_N})-(\ref{ints_Delta3}) like Eq.~(\ref{Seff}). Then, we perform analytic continuation to that equation by substituting $f_{a, n\pm} = if^{\rm R}_{a,i\omega_n\pm}$, where $a = 1,3$, $\omega_{n\pm} = \omega_n\pm ih$, and $\omega_n = (2n+1) \pi T~(n\in \mathbb{Z})$, with using
\begin{equation}
\lim{\mathcal T}{\infty}\frac{2\pi}{\mathcal T}\sum_{\eps} F_{\eps} f^{\rm R}_{\eps\pm} = 4\pi iT\sum_{n>0}f_n.
\end{equation}
We then reach the modified Usadel equation at the $\Delta_{\rm cl}$-order as $\mathcal{S}_{\mathrm{source},1}=\mathcal{S}_{\mathrm{kernel},1}[f_{1n\pm}]$, where $\mathcal{S}_{\mathrm{source},1}=1$ and

\begin{align}
\mathcal{S}_{\mathrm{kernel},1} =   \omega_{n\mp} f_{1n\pm} &- D \lr{ \alpha^2 f_{1n \mp} - [\alpha^2 + 2A_y^2 \pm 4i\alpha^3\beta A_y]  f_{1n \pm}  }     \notag \\
&-  2 \pi\nu D^2 A_y^2 \big[  K^{\rm s\pm}_{\mathrm{c},nn}(A_y)(    \beta^2\alpha^6 \! + \! A^2_y   ) f_{1n \pm} \! - \!  K^{\rm a}_{\mathrm{c},nn}(A_y) (\beta^2 \alpha^6 \! - \! A^2_y  ) f_{1n \mp} \big] \notag  \\
&+ \frac{\pi T \Gamma_{\rm s} }{8\mathcal{A}}  \sum_{\bm{q}} \sum_{n'>0}  \big[ K^{\rm s\pm}_{\mathrm{c},nn'}(\bm{q})  f_{1n' \pm} + \big( K^{\rm a}_{\mathrm{c},nn'}(\bm{q}) \! + \! K^+_{\mathrm{d},nn'}(\bm{q}) \! + \! K^-_{\mathrm{d},nn'}(\bm{q}) \big) f_{1n' \mp}    \big].     \label{Usadel1d}
\end{align}
The equation for the $\Delta^3_{\rm cl}$-order is given as $\mathcal{S}_{\rm kernel,3}[f_{3n\pm}] = \mathcal{S}_{\rm source,3}[f_{1n\pm},f_{3n\pm}]$, where
\begin{align}
\mathcal{S}_{\rm kernel,3} = \omega_{n\mp} f_{3n\pm} &- D \big(  \alpha^2 f_{3n \mp} -  [\alpha^2 + 4A^2_y \pm 4i\beta\alpha^3 A_y]f_{3n \pm}  \big) + \frac{\pi\nu D^2}{16}  \big(   \Lambda^{\pm}_{1n} f_{3n \mp} + \Lambda^{\pm}_{6n} f_{3n \pm} \big) \notag \\
&+ \frac{\pi T\Gamma_{\rm s}}{8\mathcal{A}}\sum_{\bm{q}} \sum_{n>0}  \big[K^{\rm s\pm}_{\mathrm{c},nn'}(\bm{q}) f_{3n'\pm} + (K^{\rm a}_{\mathrm{c},nn'}(\bm{q}) \!+ \! K^{+}_{\mathrm{d},nn'}(\bm{q}) \!+ \!K^{-}_{\mathrm{d},nn'}(\bm{q}))f_{3n'\mp}\big], \label{Usadel3dk}
\end{align}
and
\begin{align}
 \mathcal{S}_{\rm source,3} &=  -\frac{1}{2}[f_{1n\pm}]^2 + \frac{1}{6}\omega_{n\mp}[f_{1n\pm}]^3  + \frac{\pi\nu D}{6}  \Big( 3\alpha^2 f_{1n \pm}[f_{1n \mp}]^2 - 3\alpha^2 \,[ 1 \pm 6i\beta\alpha A_y] [f_{1n \pm}]^2f_{1n \mp} \notag \\
 &\qquad  - \alpha^2 [1 \mp 6 i \beta\alpha A_y] [f_{1n \mp}]^3 +  [\alpha^2 + 8A^2_y \pm 16i\beta\alpha^3A_y][f_{1n \pm}]^3 \Big)  \notag \\
 & \quad +  \frac{\pi\nu D^2}{16}  \Big[   \Lambda^{\pm}_{3n} f_{1n\pm} [f_{1n \mp}]^2  + \Lambda^{\pm}_{4n} [f_{1n\pm}]^2 f_{1n \mp} + \Lambda^{\pm}_{2n} [f_{1n \mp}]^3 + \Lambda^{\pm}_{5n} [f_{1n \pm}]^3 \Big]    \notag \\
 &\quad  - \frac{\pi T\Gamma_{\rm s}}{192 \mathcal{A}} \sum_{\bm{q}}\sum_{n'>0}  \Big\{ K^{\rm a}_{\mathrm{c},nn'} \big (  - [f_{1n' \mp}]^3 - 3 [f_{1n \pm}]^2 f_{1n' \mp} + 6 f_{1n \pm} f_{1n' \pm} f_{1n'\mp}   - 3 [f_{1n \mp}]^2f_{1n' \mp}      \notag \\
&\qquad - 3 f_{1n' \mp}  [f_{1n' \pm}]^2 + 3 f_{1n \mp} [f_{1n' \pm}]^2 - 6 f_{1n \mp} f_{1n \pm} f_{1n' \pm} + 3 f_{1n \mp}[f_{1n' \mp}]^2 \big) \notag\\
&\qquad + 4 K^{\rm s \pm}_{\mathrm{c},nn'}  \big( -  [f_{1n' \pm}]^3 +  3 f_{1n \pm}[f_{1n' \pm}]^2 - 3 [f_{1n \pm}]^2 f_{1n' \pm } \big) \notag \\
&\qquad - K^-_{\mathrm{d},nn'} \big( 3 f_{1n' \mp} [f_{1n'\pm}]^2 + 6  f_{1n \pm}[f_{1n' \pm}]^2  + 6 f_{1n \mp} f_{1n' \mp} f_{1n' \pm} + 6 f_{1n \pm} f_{1n \mp} f_{1n' \pm} + 3 [f_{1n  \mp}]^2 f_{1n' \mp} \big) \notag \\
&\qquad  -  \big( 4 K^+_{\mathrm{d},nn'} + K^-_{\mathrm{d},nn'} \big) \big([f_{1n' \mp}]^3  + 3\, [f_{1n \pm}]^2 f_{1n' \mp}   \big) - 12 K^+_{\mathrm{d},nn'} f_{1n \pm} [f_{1n' \mp}]^2 \Big\} \notag \\
&\quad - \frac{ \pi T \Gamma_{\rm c}}{32 \mathcal{A}} \sum_{\bm{q}}\sum_{n'>0}  \Big\{ -2 K^{\rm a}_{\mathrm{c}, nn'} \big( f_{1n\pm}  [f_{1n' \mp}]^2 +  f_{1n\mp} f_{1n'\mp} f_{1n'\pm}  \big) - 4 K^{\rm s\pm}_{\mathrm{c},nn'} f_{1n\pm}[f_{1n'\pm}]^2  \notag \\
&\qquad + K^-_{\mathrm{d},nn'} \big(   f_{1n \mp}[f_{1n' \pm}]^2 + 2 f_{1n \pm}f_{1n' \mp}f_{1n' \pm} + f_{1n \mp} [f_{1n' \mp}]^2  -  f_{1n \mp} [f_{1n' \pm}]^2 \big) \notag \\
&\qquad +  4 K^+_{\mathrm{d},nn'}  f_{1n \pm} [f_{1n' \mp}]^2    \Big\}. \label{soc3}
 \end{align}
Here, $K^{\rm s\pm}_{\mathrm{c},nn'}(\bm{q}) = K^{\rm s\pm}_{\mathrm{cR},i\omega_n,i\omega'_n} (\bm{q}),~K^{\rm a}_{\mathrm{c},nn'}(\bm{q}) = K^{\rm a}_{\mathrm{cR},i\omega_n,i\omega'_n} (\bm{q})$, $K^{\pm}_{\mathrm{d},nn'} (\bm{q}) = K^{\pm}_{\mathrm{d},i\omega_n,-i\omega'_n} (\bm{q})$, $\Lambda^{\pm}_{in}=\Lambda^{\mathrm{R}\pm}_{i\omega_n} (i=1,\dots,6)$, and we abbreviated the momentum dependence of propagators $K$ in the interaction terms of Eq.~(\ref{soc3}). Note that the momentum summation $\sum_{\bm{q}}$ is replaced with the momentum integration $\frac{\mathcal{A}}{(2\pi)^2} \int d\bm{q}$ in the calculations. In this work, the momenta are bounded within a circle with radius $q_{\rm max}=1/l_{\rm F}=1/(v_{\rm F}\tau)$.

\subsection{The simplest case}

In this section, we demonstrate how closely our modified Usadel equation in the simplest case agrees with the existing result. We consider the case where $\alpha=h=A_i=0$. This results in $f_{an+}=f_{an-}$ and
\begin{align}
K^{\rm a}_{\mathrm{c},nn'}(\bm{q})=0,&\quad K^{-}_{\mathrm{d},nn'}(\bm{q})=0, \\
K^{\rm s\pm}_{\mathrm{c},nn'}(\bm{q})=\frac{4}{\pi\nu}\mathcal{D}_{nn'}(\bm{q}),&\quad K^{+}_{\mathrm{d},nn'}(\bm{q})=\frac{4}{\pi\nu}\mathcal{D}_{nn'}(\bm{q}),
\end{align}
where $\mathcal{D}^{-1}_{nn'}(\bm{q}) = D\bm{q}^2 +\omega_n + \omega'_{n}$. In this way, the modified Usadel equation at the $\Delta_{\rm cl}$-order becomes
\begin{equation}
  \omega_n f_{1n} + \frac{T \Gamma_{\rm s}}{\nu}\sum_{n'>0}\int \frac{d\bm{q}}{(2\pi)^2} \mathcal{D}_{nn'}(\bm{q}) f_{1n'} = 1. \label{sudel1}
\end{equation}
Note that the term from the localization correction due to the Cooper channel vanishes in this equation. This fact emphasizes the minor role of the Cooper interaction compared to the Coulomb interaction. In the same way, the modified Usadel equation at the $\Delta^3_{\rm cl}$-order becomes
\begin{align}
\omega_n f_{3n} &+ \frac{T \Gamma_{\rm s}}{\nu}\sum_{n'>0}\int \frac{d\bm{q}}{(2\pi)^2} \mathcal{D}_{nn'}(\bm{q}) f_{3n'} \notag \\
&= -\frac{1}{2}[f_{1n}]^2 + \frac{1}{6}\omega_n[f_{1n}]^3 + \frac{T\Gamma_{\rm s}}{6\nu} \sum_{n'>0} \int \frac{d\bm{q}}{(2\pi)^2} \mathcal{D}_{nn'}(\bm{q}) \lrb{ [f_{1n'}]^3 + 3[f_{1n}]^2f_{1n'} }  . \label{sudel3}
\end{align}
By some calculations, we can show that Eqs.~(\ref{sudel1}) and (\ref{sudel3}) can be achieved by expanding $\theta_{n}=\Delta_{\rm cl}f_{1n}+\Delta^3_{\rm cl}f_{3n}$ in the following equation to the $\Delta^3_{\rm cl}$-order:
\begin{align}
\omega_n\sin\theta_n - \Delta\cos\theta_n + \frac{T\Gamma_{\rm s}}{\nu} \cos\theta_n \sum_{n'>0}\int \frac{d\bm{q}}{(2\pi)^2}   \frac{\sin\theta_{n'}}{D\bm{q}^2+\omega_n+\omega'_n} = 0. \label{sudel}
\end{align}
On the other hand, by writing in our manner, the existing result \cite{Andriyakhina2} is, 
\begin{equation}
\omega_n\sin\theta_n - \Delta\cos\theta_n + \frac{4\pi T \Gamma_{\rm s} D}{g} \cos\theta_n \sum_{n'>0}\int \frac{d\bm{q}}{(2\pi)^2} \frac{\sin\theta_{n'}}{D\bm{q}^2 + \mathcal{E}_n + \mathcal{E}_n'} = 0, \label{ude}
\end{equation}
where $g$ is dimensionless conductance, which is already defined in the text at the top of Sec. \ref{Numerical}, and $\mathcal{E}_n \equiv \omega_n\cos\theta_n + \Delta_{\rm cl}\sin\theta_n$. To compare this with Eq.~(\ref{sudel}), first note that in Ref.~\cite{Andriyakhina2}, $g=2\pi \nu_{\rm F} D$, where $\nu_{\rm F}$ is the density of states. In our case, $\nu_{\rm F} = 2\nu$. Thus, $g=4\pi\nu D$, and the coefficients in front of the integrals in Eqs.~(\ref{sudel}) and (\ref{ude}) are equivalent. In this way, equations (\ref{sudel}) and (\ref{ude}) differ only in the denominator of the integrand. Indeed, they are the same at the $\Delta_{\rm cl}$-order but become different at the $\Delta^3_{\rm cl}$-order. This controversy occurs because we considered the action $S_{\rm II}$, which contains a contribution like $\mathrm{Tr}\,\{WM_1WM_2\}$ for certain $M_1$ and $M_2$, as a perturbative term, and it does not contribute at the first cumulant expansion in Eq.~(\ref{Seff}).

\end{widetext}

The problem can be fixed by two approaches. One is to expand the perturbation theory to a higher order. At the second cumulant, the term like $\braket{S_{\rm II}S_{\rm int}}_0$ can give a nonzero contribution. The other one is to include $S_{\rm II}$ into the free action $S_0$ in the first place. Indeed, we can also include the interaction action $S_{\rm int}$ into $S_0$. In that case, there are additional contributions from higher orders of $\Gamma_{\rm c}$ and $\Gamma_{\rm s}$ to the modified Usadel equation \cite{Andriyakhina2}.

However, when $\alpha,~h$, and $A_i$ are nonzero, the calculations in both approaches mentioned above are not easy. Nevertheless, these calculations, even in this simplest case, are accompanied by the one-loop integration of the terms like $[\mathcal{D}_{nn'}(\bm{q})f_{an'}]^2$, which are not important at high temperature. First, the integration of $[\mathcal{D}_{nn'}(\bm{q})f_{an'}]^2$ over $\bm{q}$ becomes finite if the momentum bound is brought to infinity and the temperature is not zero. This is different from the terms like $\mathcal{D}_{nn'}(\bm{q})f_{an'}$ which are taken into account in this work. Those terms logarithmically diverge and need a momentum bound like the one that we have described at the end of the previous section. Moreover, since $f_{1n}\sim 1/\omega_n$ and $f_{3n}\sim 1/\omega^3_n$, the ratio between the one-loop integration of $[\mathcal{D}_{nn'}(\bm{q})f_{an'}]^2$ and that of $\mathcal{D}_{nn'}(\bm{q})f_{an'}$ is proportional to $1/\omega_n$ or $1/\omega^3_n$, and thus, the terms such as $[\mathcal{D}_{nn'}(\bm{q})f_{an'}]^2$ become important near the absolute zero temperature. Therefore, the calculation including only one-loop integration of one propagator is valid at finite temperatures.

\subsection{Criteria for neglecting the triplet interaction in any case}
\label{AppendixD5}

In this section, we discuss the criteria for definitely neglecting the triplet channel of interaction. It is argued in Refs.~\cite{Finkelstein1, Burmistrov11} that the triplet channel can be neglected from the renormalization group (RG) equation if the triplet diffusive modes acquire sufficiently large mass. Since the perturbative method in Ref.~\cite{Andriyakhina2}, which is similar to ours, gives qualitatively the same results as the RG method, the criteria due to the RG method should also be rendered to justify whether the triplet channel can be neglected in our case.

We begin with the criterion in Ref.~\cite{Finkelstein1}: The triplet channel can be neglected if $\Gamma_{\rm r}\gg T_{\rm c0}$. With the numerical parameters specified in Sec. \ref{Numerical}, the comparable SOC case satisfies this condition, whereas the weak SOC case does not. Next, we look at the criterion in Ref.~\cite{Burmistrov11}: The triplet channel can be neglected if $L\gg L_{\rm DP}=v_{\rm F}/\Delta_{\rm DP}$, where $\Gamma_{\rm r} = \Delta^2_{\rm DP}\tau$ and $L$ is the RG characteristic length scale. Following Ref.~\cite{Andriyakhina2}, within our perturbative approach, this $L$ corresponds to the length at the upper bound of the triplet propagator when neglecting $A_i,h$, and $q_i$: $L$ may be approximated by the smearing thermal length $L_T = \sqrt{D/T}$. The comparable SOC case satisfies again the condition $L_T > L_{\rm DP}$ at $T<T_{\rm c}$, while the weak SOC case does not satisfy the condition in the interval $T\in[T^*_{\rm c},T_{\rm c}]$.

Although the parameters for the weak SOC case do not satisfy the two criteria discussed above, this does not mean that the triplet channel is definitely important. In perturbative methods like ours, which are valid if the disorder strength is weak (see the text at the top of Sec. \ref{Numerical}), only the weakness of the bare triplet interaction is required. Finding and analyzing a concrete model that satisfies this condition is considered future work.

\section*{Appendix E: Cooper pair momenta of zero-current superconducting states}
\label{AppendixE}
\renewcommand{\theequation}{E.\arabic{equation}}

\setcounter{equation}{0}
\setcounter{subsection}{0}
\def\thesubsection{\arabic{subsection}}

\begin{figure}[t]
\begin{minipage}{0.49\hsize}
\subfloat[\label{fig1as}]{\includegraphics[width=4cm]{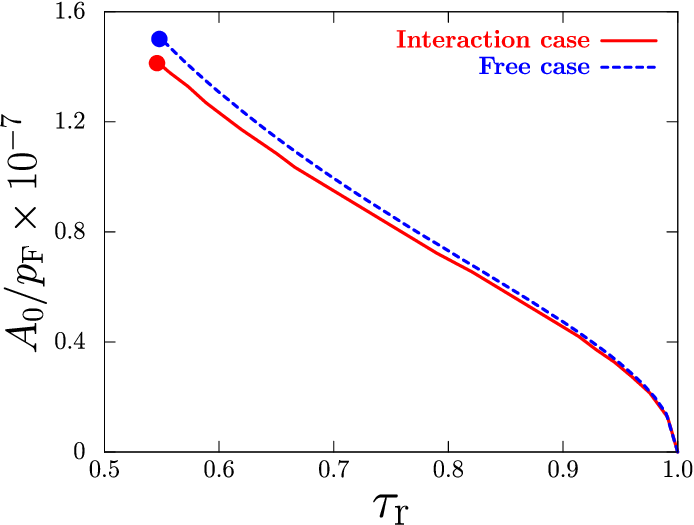}}
\end{minipage}
\begin{minipage}{0.49\hsize}
\subfloat[\label{fig1bs}]{\includegraphics[width=4cm]{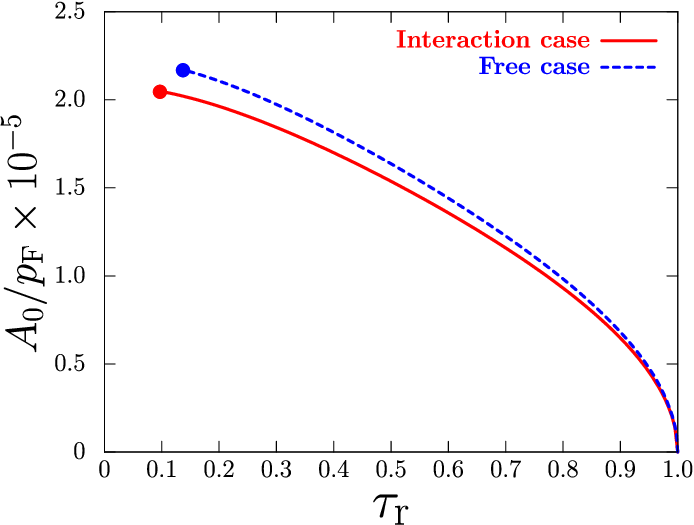}}
\end{minipage}

\caption{The gauge field $A_0$ corresponding to the momentum of Cooper pairs for (a) weak SOC case and (b) comparable SOC case. We plot the normalized temperature $\tau_{\rm r}$ dependence by varying the magnetic field and temperature on the SC transition line, $T=T_{\rm c}(h)$ (with interactions) or $T=T_{\rm c0}(h)$ (without interactions). The red solid (blue dashed) line is obtained by taking into account (neglecting) the effects of interactions. The red (blue) circle denotes the tricritical point in the interaction (free) case.}
\label{fig1s}
\end{figure}

We show the gauge field $A_0$, namely the Cooper pair momenta in zero-current superconducting (SC) states, at each point on the SC transition line. The results for the weak SOC and comparable SOC cases, which are defined in Sec. \ref{Numerical}, are presented in Figs.~\ref{fig1as} and \ref{fig1bs}, respectively. The gauge field is slightly suppressed by the effects of interactions in both cases, indicating that the ordering of helical SC states is hindered by the Coulomb interaction. Nevertheless, as described in Sec. \ref{NumericalB}, the SD effect in the high-normalized-temperature regime is unchanged. Interestingly, the $A_0$ curve in the comparable SOC case is concave, while the $A_0$ curve in the weak SOC case becomes convex near the tricritical point.

\section*{Appendix F: Zero-field transition temperatures} 
\label{AppendixF}
\renewcommand{\theequation}{F.\arabic{equation}}

\setcounter{equation}{0}
\setcounter{subsection}{0}
\def\thesubsection{\arabic{subsection}}

\begin{figure}[t]
\begin{minipage}{0.49\hsize}
\subfloat[\label{fig2as}]{\includegraphics[width=4.0cm]{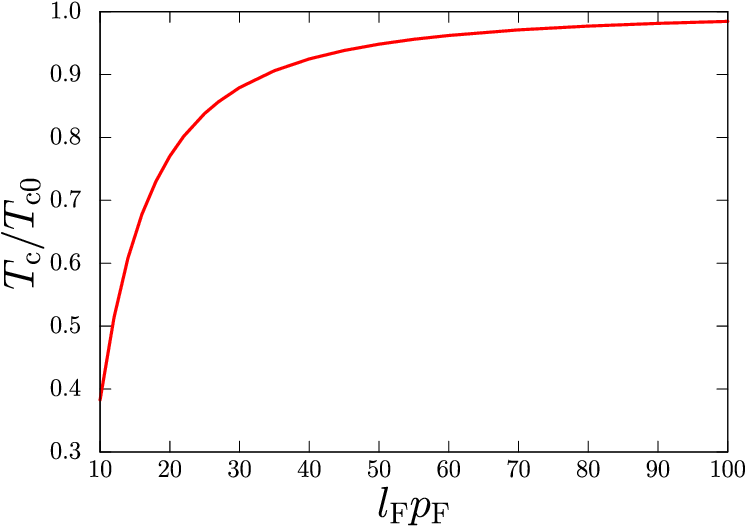}}
\end{minipage}
\begin{minipage}{0.49\hsize}
\subfloat[\label{fig2bs}]{\includegraphics[width=4.0cm]{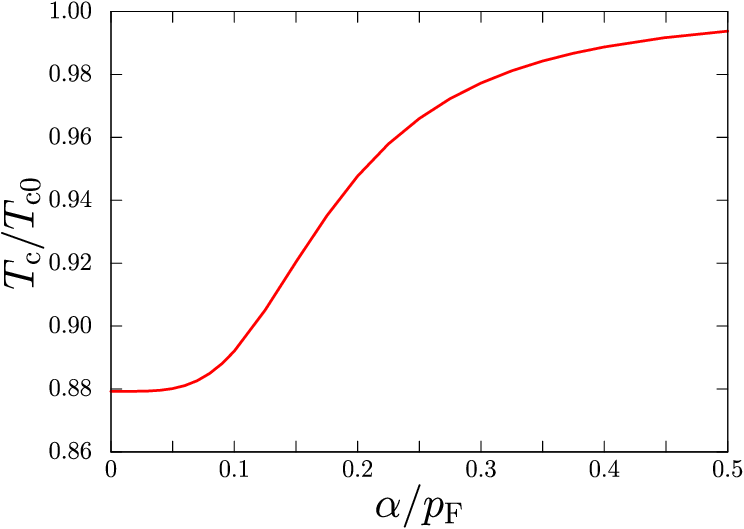}}
\end{minipage}
\begin{minipage}{0.49\hsize}
\subfloat[\label{fig2cs}]{\includegraphics[width=4.0cm]{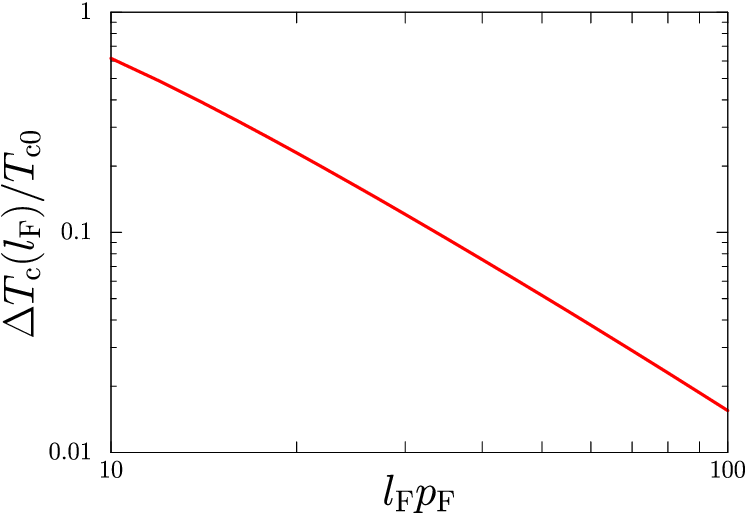}}
\end{minipage}
\begin{minipage}{0.49\hsize}
\subfloat[\label{fig2ds}]{\includegraphics[width=4.0cm]{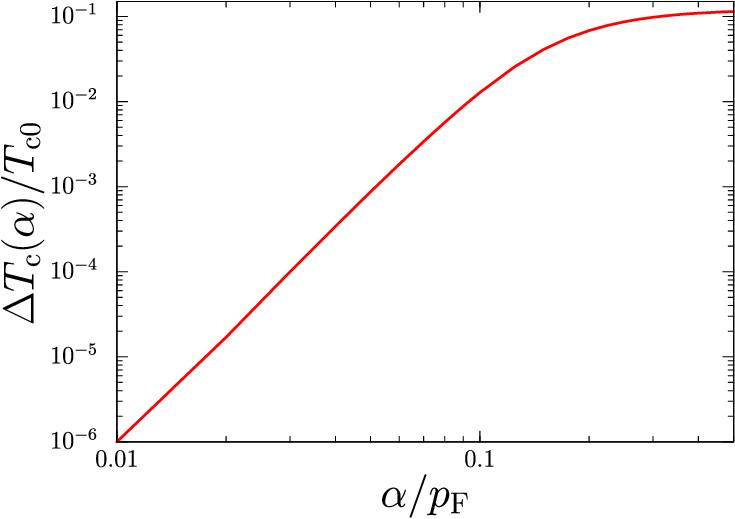}}
\end{minipage}
\caption{The zero-field SC transition temperature $T_{\rm c}$ in the interaction case. (a) The mean field path $l_{\rm F} = v_{\rm F}\tau$ dependence for the Rashba SOC $\alpha=0.015 \, p_{\rm F}$. (b) The Rashba SOC dependence for the mean free path $l_{\rm F}=30$. (c) and (d) Logarithmic plot for $\Delta T_{\rm c}(l_{\rm F})=T_{\rm c}(l_{\rm F}=\infty) - T_{\rm c}(l_{\rm F})$ and $\Delta T_{\rm c}(\alpha) = T_{\rm c} (\alpha)-T_{\rm c}(\alpha=0)$.}

\label{fig2s}
\end{figure}

We show the disorder dependence of the zero-field SC transition temperature $T_{\rm c}$ in the interacting system. Figure~\ref{fig2as} presents the mean free path $l_{\rm F} = v_{\rm F}\tau$ dependence of $T_{\rm c}$ for the Rashba SOC $\alpha=0.015 \, p_{\rm F}$, while Fig.~\ref{fig2as} shows the Rashba SOC dependence for $l_{\rm F}=30$. As the disorder strength reduces the mean free path, the zero-field transition temperature decreases due to the cooperation of localization effects and interactions [Fig.~\ref{fig2as}]. However, the Rashba SOC decreases suppression of $T_{\rm c}$ [Fig.~\ref{fig2bs}]. Note that $T_{\rm c}(l_{\rm F}=\infty) = T_{\rm c0}$ because the momentum integration bound $q_{\rm max}$ in the modified Usadel equation becomes zero when $l_{\rm F}\rightarrow \infty$ [see the text below Eq. (\ref{soc3})]. We also show a logarithmic plot in Figs.~\ref{fig2cs} and \ref{fig2ds}, where $\Delta T_{\rm c}(l_{\rm F})=T_{\rm c}(l_{\rm F}=\infty) - T_{\rm c}(l_{\rm F})$ and $\Delta T_{\rm c}(\alpha) = T_{\rm c} (\alpha)-T_{\rm c}(\alpha=0)$ are depicted, respectively. When we define the exponents $u$ and $v$ as $\Delta T_{\rm c}(l_{\rm F})\propto l_{\rm F}^{u}$ and $\Delta T_{\rm c}(\alpha)\propto \alpha^v$, the exponents change as $l_{\rm F}$ and $\alpha$ are varied. We obtain $u\sim-1.8$ and $v\sim 4.1$ in the long $l_{\rm F}$ and small $\alpha$ region, respectively.

\section*{Weak localization conductivity}
\label{AppendixG}
\renewcommand{\theequation}{G.\arabic{equation}}

\setcounter{equation}{0}
\setcounter{subsection}{0}
\def\thesubsection{\arabic{subsection}}

\subsection{Formulation}

First, we present how the weak-localization (WL) correction to conductivity, namely the WL conductivity $\sigma_{\rm WL}$, is calculated using the Keldysh functional formalism. This WL conductivity only involves cooperon modes at zero frequency and does not include the effect of e-e interactions, as seen in Ref.~\cite{Burmistrov3}. It is convenient to work with the normal saddle point $\bar{Q}'_{\rm N}=U'\Lambda'_{\rm N}U'^{-1}$, where $\Lambda'_{\rm N}=\mathrm{diag}\,\{\gamma_3,\gamma_3\}$ and  $U'=\mathrm{diag}\,\{\check{U}_{\rm K},\check{U}^{\mathbb{T}}_{\rm K}\}$, instead of that used in Sec. \ref{Method}, $\bar{Q}_{\rm N}$. Here, $U^{\mathbb{T}}$ denotes the transpose of a matrix and the interchange of the time of $U$, which means that $U^{\mathbb{T}}_{t_1t_2}=U^{\rm t}_{t_2t_1}$ with $U^{\rm t}_{t_2t_1}$ being the ordinary transpose in the matrix elements of $U_{t_2t_1}$. The matrix $\check{U}_{\rm K}$ has the same form as Eq.~(\ref{Ukeys}), but $\tau_0$ is replaced with $\sigma_0$. This saddle point $\bar{Q}'_{\rm N}$ comes from the time-reversal symmetry condition of the matrix field $Q$, namely $Q^{\mathbb{T}}=-CQC$ with $C=\tau_2\gamma_0\sigma_2$. To the best of our knowledge, this saddle point $\bar{Q}'_{\rm N}$ was first introduced in Ref.~\cite{Kamenevhon} to calculate the WL conductivity in the spinless case within the Keldysh functional formalism. When calculating $\sigma_{\rm WL}$ at the lowest correction of disorder, only the cooperon modes are needed. The choice of saddle point gives us a structure of the fluctuation matrix $W$ as follows:
\begin{align}
W=&\begin{pmatrix}
0 & \check{C}_2 \\
\check{C}_1 & 0
\end{pmatrix},\\ 
\check{C}_1 = \begin{pmatrix}
0 & \sigma_2 \hat{C}^{\mathbb{T}}_{1}\sigma_2 \\
\hat{C}_1 & 0
\end{pmatrix}&,\quad \check{C}_2 = \begin{pmatrix}
0 & \hat{C}_2 \\
 \sigma_2 \hat{C}^{\mathbb{T}}_{2}\sigma_2 & 0
\end{pmatrix}.
\end{align}
Since the saddle point chosen here is different from that adopted in the formulation of superconducting states, several operators in the NLSM action need to be modified. Since we neglect the interaction terms of action in the present approximation, let us look at the changes in the non-interacting part, $S_0$. We find that the order of the multiplication of the Pauli matrices in Keldysh and Nambu spaces ($\gamma_i$ and $\tau_i$, respectively) should be interchanged. Explicitly, 

\begin{widetext}
\begin{align}
\gamma_0\check{E}=\gamma_0\tau_3(\eps+h\sigma_3)&\quad \rightarrow \quad E'=\tau_3\gamma_0(\eps+h\sigma_3), \\
\gamma_0\check{A}_x=\gamma_0\check{L}_{\rm r}(\tau_3A_x+\tau_0\sigma_2\alpha)\check{L}_{\rm r} &\quad \rightarrow \quad A'_x=\tau_3\check{L}_{\rm r}(\gamma_0A_x+\tau_3\gamma_0\sigma_2\alpha)\check{L}_{\rm r},         \\
\gamma_0\check{A}_x=\gamma_0\check{L}_{\rm r}(\tau_3A_y-\tau_0\sigma_1\alpha)\check{L}_{\rm r} &\quad\rightarrow \quad A'_y=\tau_3\check{L}_{\rm r}(\gamma_0A_y - \tau_3\gamma_0\sigma_1\alpha)\check{L}_{\rm r},          \\
\gamma_0\check{\nabla}_i=\gamma_0\tau_0\nabla_i-i[\check{A}_i,\dots] & \quad \rightarrow \quad \nabla'_i = \tau_0 \gamma_0 \nabla - i[A'_i,\dots], \\
\gamma_0\check{F}_{ij} = \gamma_0(\partial_i\check{A}_j - \partial_j\check{A}_i - i[\check{A}_i,\check{A}_j]) &\quad \rightarrow \quad F'_{ij} =  \partial_iA'_j - \partial_jA'_i - i[A'_i,A'_j].
\end{align}
These modifications, together with the matrix $C=\tau_2\gamma_0\sigma_2$, are considered straightforward extensions of the spinless case in the previous work \cite{Kamenevhon}. Following the procedure in Sec.~\ref{MethodB} and using the choice
\begin{align}
\hat{C}_1 = c_{10} + ic_{11}\sigma_1 + c_{12}\sigma_2 + c_{13}\sigma_3, \quad \hat{C}_2 = -c_{20} + ic_{21}\sigma_1 - c_{22}\sigma_2 - c_{23}\sigma_3, 
\end{align}
we reach the action of the cooperon modes as follows:
\begin{align}
&S_0 = \frac{i\pi\nu}{2} \sum_{\eps,\eps'} \Phi^{(2) \dagger}_{\eps\eps'}(-\bm{q}) \,  \mathcal{D}^{-1}_{\rm c}(\tilde{\bm{q}},\eps+\eps') \, \Phi^{(1)}_{\eps'\eps} (\bm{q}), \\
&\Phi^{(i)}_{\eps\eps'}(\bm{q}) = (c_{i0,\eps\eps'}(\bm{q}), c_{i1,\eps\eps'}(\bm{q}),   c_{i2,\eps\eps'}(\bm{q}),  c_{i3,\eps\eps'}(\bm{q})  )^{\rm t},
\end{align}
where $\mathcal{D}^{-1}_{c}$ is exactly the same as the retarded propagator in Eq.~(\ref{Cooperon}). We have used $(c^{\mathbb{T}}_{ij})_{\eps\eps'} = c_{ij,-\eps',-\eps}$ to derive the above equation. In this way, the propagators of cooperon modes are given as follows:
\begin{equation}
\braket{c_{1i\eps_1\eps_2}(\bm{q}_1)c_{2j\eps_3\eps_4}(\bm{q}_2)}_0 = \frac{2}{\pi\nu} \delta_{\bm{q}_1,-\bm{q}_2}  \delta_{\eps_1\eps_4}\delta_{\eps_2\eps_3} [\mathcal{D}_{\rm c}(\tilde{\bm{q}}_1,\eps_1+\eps_2)]_{ij}.
\end{equation}
\end{widetext}

To calculate conductivity in the present formalism, we have to add the quantum part of the gauge field to the action \cite{Kamenev1,Kamenevhon}. This can be done by replacing $\gamma_0A_i$ with $\gamma_0A_{\mathrm{cl}i}+\gamma_1A_{\mathrm{q}i}$, where $A_{\mathrm{cl}i}$ and $A_{\mathrm{q}i}$ denote the classical and quantum parts of the gauge field, respectively. The conductivity is then calculated using the following equation:
\begin{equation}
\sigma(\omega) = -\frac{1}{2\omega} \bigg[      \frac{\delta^2Z}{\delta A_{\mathrm{cl}i}(\omega)\delta A_{\mathrm{q}i}(-\omega)} \bigg]_{A=0},
\end{equation}
where $Z=\int DQ e^{iS'_0}$ and $S'_0 = S'_0[W;\bm{A}_{\rm cl},\bm{A}_{\rm q}]$ is the non-interacting action including the quantum part of the gauge field. By dividing the action into the gauge-field-independent part and the part that is proportional to $A_{\mathrm{cl}i}A_{\mathrm{q}i}$, namely $S'_0[W] = S_0[W] + \delta S[W;\bm{A}_{\rm cl},\bm{A}_{\rm q}]$, we have
\begin{align}
\sigma(\omega) &= -\frac{i}{2\omega} \bigg[      \frac{\delta^2   \big\langle \delta S[W;\bm{A}_{\rm cl},\bm{A}_{\rm q}]\big\rangle_0       }{\delta A_{\mathrm{cl}i}(\omega)\delta A_{\mathrm{q}i}(-\omega)}      \bigg]_{A=0}, \\
\delta S &= -\frac{i\pi\nu D}{4}\mathrm{Tr}\,\big\{  [\overleftrightarrow{A}_{\mathrm{cl}i},\Lambda'_{\rm N}] \, [\overleftrightarrow{A}_{\mathrm{q}i},\Lambda'_{\rm N}] \notag \\
&\quad + \frac{1}{4}   [\overleftrightarrow{A}_{\mathrm{cl}i},[W,\Lambda'_{\rm N}]] \, [\overleftrightarrow{A}_{\mathrm{q}i},[W,\Lambda'_{\rm N}]]     \big\},  \label{Sdw}
\end{align}
where $\overleftrightarrow{A}_{\mathrm{cl}i}=U'\tau_3\gamma_0U' A_{\mathrm{cl}i}$ and $\overleftrightarrow{A}_{\mathrm{q}i}=U'\tau_3\gamma_1U' A_{\mathrm{q}i}$. The first term in the curly bracket of the r.h.s. of Eq.~(\ref{Sdw}) is responsible for the Drude conductivity, while the other is the WL correction. The terms involving the gradient terms of $W$ are neglected for simplicity. The terms associated with the topological action also vanish in the present approximation because we are only interested in the diagonal $ii$-component of conductivity. With the help of the formula  $\int_{-\infty}^{\infty} d\eps \, (F_{\eps+\omega}-F_{\eps}) = 2\omega $, we obtain the equation for the weak localization correction to conductivity $\sigma_{\rm WL}$ in the DC limit $(\omega\rightarrow0)$ as follows:
\begin{equation}
\sigma_{\rm WL}/\sigma_{\rm D} = \frac{1}{4\mathcal{A}}\sum_{\bm{q}}\mathrm{tr}\,\{ \Pi \mathcal{D}_{\rm c}(\bm{q},\omega=0) \},
\end{equation}
where $\sigma_{\rm D}=2\nu D$ and $\Pi=\mathrm{diag}\,\{1,-1,-1,-1\}$. This result is consistent with previous work \cite{Kashuba, Marinescu, ilic2, Burmistrov3}. In the absence of SOC, $\sigma_{\rm WL}$ becomes independent of the in-plane Zeeman field, as argued explicitly in \cite{ilic2}. The above equation has an ultraviolet divergence and can also have an infrared divergence if $\alpha=0$. In the numerical analysis, we choose $q_{\rm max}=1/L_T$ and $q_{\rm min}=1/L_{\phi}$ as the upper and lower bounds of the momentum integration, respectively, where $L_T=\sqrt{D/T}$ is the thermal smearing length and $L_{\phi}$ is the dephasing length. The latter is given here as $L_{\phi}p_{\rm F}=10^6$.

\begin{figure}[b]

\includegraphics[width=8.5cm]{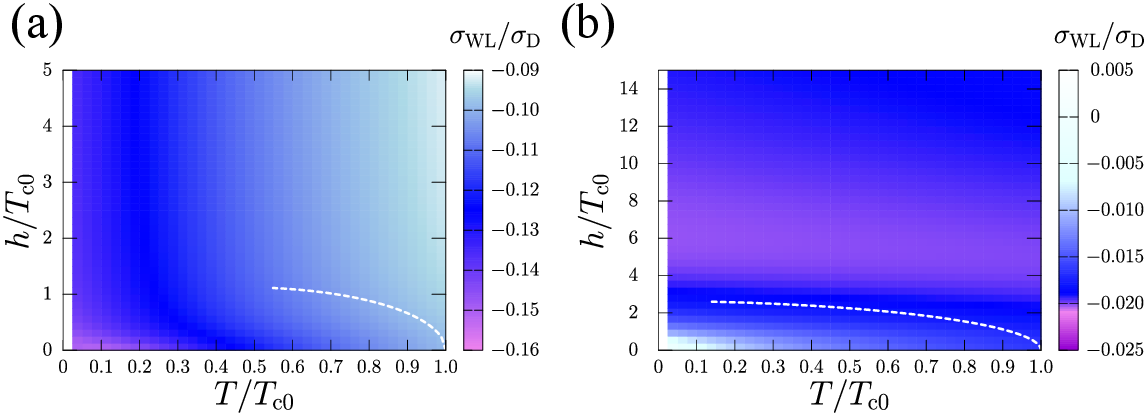}

\label{fig3s}

\caption{The color map of the WL correction to conductivity $\sigma_{\rm WL}$ including the extremely-high-field regime for (a) weak SOC case and (b) comparable SOC case defined in Sec. \ref{Numerical}. The dashed white curves show the SC transition lines. }

\end{figure}

\subsection{Numerical results in the extremely-high-field regime}

In Sec. \ref{NumericalC}, the WL correction to conductivity $\sigma_{\rm WL}$ is calculated in the region where the SC states are stable. Here, we show the results in extremely high fields for the weak SOC and comparable SOC cases in Figs.~\hyperref[fig3s]{6a} and \hyperref[fig3s]{6b}, respectively. As shown in Fig.~\hyperref[fig3s]{6b}, at $h\gtrsim 5T_{\rm c0}$, the magnitudes of $\sigma_{\rm WL}$ in the comparable SOC case decrease as the temperature and magnetic field increase. The same behavior is found in almost the whole region of the weak SOC case, in which the trend does not change even in the extremely-high-field regime, as presented in Fig.~\hyperref[fig3s]{6a}. This is in contrast to the comparable SOC case, in which the magnitudes of $\sigma_{\rm WL}$ increase as the temperature and magnetic field increase at $h\lesssim5T_{\rm c0}$. Note also that $\sigma_{\rm WL}$ becomes a positive value at sufficiently low fields.

\end{document}